%% file: FSQ-16-007_temp.tex
\pdfoutput=1

\documentclass[11pt,twoside,a4paper,cmspaper,final,collab]{cms-tdr}
\def\svnVersion{32575ea-D}\def\svnDate{2019/07/23}\def\cmsCernNoTag{CERN-EP-2018-285}\def\cmsCernDate{\today}\def\cmsMessage{"Published in the European Physical Journal C as \href{http://dx.doi.org/10.1140/epjc/s10052-019-7202-9}{\doi{10.1140/epjc/s10052-019-7202-9}.}"}

\begin{document}\cmsNoteHeader{FSQ-16-007}

\hyphenation{had-ron-i-za-tion}
\hyphenation{cal-or-i-me-ter}
\hyphenation{de-vices}
\RCS$HeadURL$
\RCS$Id$
\newlength\cmsFigWidth
\newlength\cmsTabSkip\setlength{\cmsTabSkip}{1ex}
\ifthenelse{\boolean{cms@external}}{\setlength\cmsFigWidth{0.85\columnwidth}}{\setlength\cmsFigWidth{0.4\textwidth}}
\ifthenelse{\boolean{cms@external}}{\providecommand{\cmsLeft}{top\xspace}}{\providecommand{\cmsLeft}{left\xspace}}
\ifthenelse{\boolean{cms@external}}{\providecommand{\cmsRight}{bottom\xspace}}{\providecommand{\cmsRight}{right\xspace}}
\providecommand{\NA}{\ensuremath{\text{---}}}
\cmsNoteHeader{FSQ-16-007}
\title{\texorpdfstring{Measurement of exclusive $\Pgr^{0}$~photoproduction in ultraperipheral pPb collisions at $\sqrtsNN = 5.02\TeV$}{Measurement of exclusive rho(770)0 photoproduction in ultra-peripheral pPb collisions at sqrt(sNN) = 5.02 TeV}}

\date{\today}

\abstract{
Exclusive $\Pgr^{0}$ photoproduction is measured for the first time in ultraperipheral \Pp{}Pb collisions at $\sqrtsNN = 5.02\TeV$ with the CMS detector. The cross section $\sigma (\Pgg \Pp\to \Pgr^{0}\Pp)$ is $11.0 \pm 1.4\stat \pm 1.0\syst$ \unit{$\mu$b} at $\langle W_{\Pgg\Pp}\rangle = 92.6\GeV$ for photon-proton centre-of-mass energies $W_{\Pgg\Pp}$ between $29$ and $213\GeV$. The differential cross section $\rd\sigma/\rd\abs{t}$ is measured in the interval $0.025 < \abs{t} < 1\GeV^{2}$ as a function of $W_{\Pgg\Pp}$, where $t$ is the squared four-momentum transfer at the proton vertex. The results are compared with previous measurements and theoretical predictions. The measured cross section $\sigma (\Pgg \Pp\to \Pgr^{0}\Pp)$ has a power-law dependence on the photon-proton centre-of-mass, consistent with electron-proton collision measurements performed at HERA. The $W_{\Pgg\Pp}$ dependence of the exponential slope of the differential cross section $\rd\sigma/\rd\abs{t}$ is also measured.
}
\hypersetup{%
pdfauthor={CMS Collaboration},%
pdftitle={Measurement of exclusive rho(770)0 photoproduction in ultra-peripheral pPb collisions at sqrt(sNN) = 5.02 TeV},%
pdfsubject={CMS},%
pdfkeywords={CMS, UPC, vector meson, exclusive photoproduction}}

\maketitle

\hypersetup{pageanchor=true}
\section{Introduction}
{\tolerance=1200
Exclusive vector meson (VM) photoproduction, $\Pgg\Pp \to \mathrm{VM}\Pp$, has received renewed interest following recent studies of ultraperipheral collisions involving ions and protons at the CERN LHC~\cite{Baltz:2007kq,Contreras:2015dqa}. In such collisions, photon-induced interactions predominantly occur when the colliding hadrons are separated by a distance larger than the sum of their radii. In this case, one of the hadrons may emit a quasi-real photon that fluctuates into a quark-antiquark pair with the quantum numbers of the photon, which can then turn into a VM upon interacting with the other hadron. The interaction of the VM with the hadron proceeds via the exchange of the vacuum quantum numbers, the so-called pomeron exchange.
 Proton-lead~(\Pp{}Pb) collisions are particularly interesting for studying photon-proton interactions~\cite{Frankfurt:2006tp,Guzey:2013taa} because the large electric charge of the Pb nucleus strongly enhances photon emission. Also, in these events, one can determine the photon direction and hence the photon-proton centre-of-mass energy $W_{\Pgg\Pp}$ unambiguously. This advantage is not present in symmetric colliding systems such as \Pp{}\Pp~interactions.
Exclusive VM photoproduction is interesting because the Fourier transform of the $t$ distribution, with $t$ being the squared four-momentum transfer at the proton vertex, is related to the two-dimensional spatial distribution of the struck partons in the plane transverse to the beam direction.
Furthermore, some models suggest that the energy dependence of the integrated cross section and that of the $t$ distribution may provide evidence of gluon saturation, as discussed in Refs.~\cite{Armesto:2014sma,Toll:2012mb,Jones:2013pga,Goncalves:2018blz,Cepila:2018zky,Cepila:2016uku}.
\par}

{\tolerance=300
By using ultraperipheral \Pp{}Pb collisions at $\sqrtsNN = 5.02\TeV$ at the LHC, the ALICE Collaboration has measured the exclusive photoproduction of \PJgy mesons in the centre-of-mass energy interval $20< W_{\Pgg\Pp} < 700\GeV$~\cite{TheALICE:2014dwa,Abelev:2012ba}.
The LHCb Collaboration has studied exclusive \PJgy, \Pgy, and \PgU(nS)~photoproduction in \Pp{}\Pp~collisions at $\sqrt{s} = 7$ and $8\TeV$~\cite{Aaij:2014iea,Aaij:2015kea}.
Exclusive photoproduction of $\Pgr^{0}$~mesons was first studied in fixed-target experiments at $W_{\Pgg\Pp}$ values up to 20\GeV~\cite{Bauer:1977iq,Crittenden:1997yz}. Experiments at the HERA electron-proton collider at DESY have studied this process at $W_{\Pgg\Pp}$ values ranging from 50 to 187\GeV, both with quasi-real photons and for photons with larger virtualities~\cite{Breitweg:1997ed,Aid:1996bs}. The HERA data have provided clear experimental evidence for the transition from the soft to the hard diffractive regime~\cite{Newman:2013ada,Favart:2015umi}.
}
{\tolerance=300
More recently, exclusive photoproduction of $\Pgr^{0}$ mesons has been studied by the STAR Collaboration in ultraperipheral AuAu collisions at the BNL RHIC collider~\cite{Adler:2002sc,Abelev:2007nb,Adamczyk:2017vfu}, and by the ALICE Collaboration in PbPb collisions~\cite{Adam:2015gsa}. The cross sections measured by the ALICE and STAR Collaborations in photon-nucleus interactions are 40\% lower than both the prediction from the Glauber approach and the corresponding measurements in photon-proton interactions~\cite{Adam:2015gsa,Frankfurt:2015cwa}. However, the Glauber approach reproduces the measured cross sections well at lower energies. This is an indication that nuclei do not behave as a collection of independent nucleons at high energies.
}
In the present analysis, exclusive photoproduction of $\Pgr^{0}$ mesons in the \Pgpp\Pgpm decay channel in ultraperipheral \Pp{}Pb collisions at $\sqrtsNN = 5.02$\TeV is measured. The cross section is measured as a function of $W_{\Pgg\Pp}$ and $t$.
In this paper $\abs{t}$ is defined as the squared transverse momentum of the $\Pgr^0$~meson, $\abs{t} \approx \pt^{2}$.

{\tolerance=300
This paper is organized as follows. Section~\ref{sec:detector} describes the experimental apparatus and Section~\ref{sec:Data} the data and simulated Monte Carlo samples. The event selection procedure is illustrated in Section~\ref{sec:sele}. Section~\ref{sec:BG} discusses the background contributions and Section~\ref{sec:minv_fit} the strategy used to extract the signal; the systematic uncertainties are summarized in Section~\ref{sec:xsec2}. The total and differential cross sections are presented in Section~\ref{sec:xsec}. The results are summarized in Section~\ref{sec:con}.
}
\section{The CMS detector}
\label{sec:detector}
{\tolerance=440
The central feature of the CMS apparatus is a superconducting solenoid of 6\unit{m} internal diameter, providing a magnetic field of 3.8\unit{T}. Within the solenoid volume are a silicon pixel and strip tracker, a lead tungsten crystal electromagnetic calorimeter (ECAL), and a brass and scintillator hadron calorimeter (HCAL), each composed of a barrel and two endcap sections. The silicon tracker measures charged particles within the range $\abs{\eta}< 2.5$. It consists of 1440 silicon pixel and 15\,148 silicon-strip detector modules and is located in the field of the superconducting solenoid. For nonisolated particles of $1 < \pt < 10\GeV$ and $\abs{\eta} < 1.4$, the track resolutions are typically 1.5\% in \pt and 25--90 (45--150)\mum in the transverse (longitudinal) direction~\cite{Chatrchyan:2014fea}.

The pseudorapidity coverage for the ECAL and HCAL detectors is $\abs{\eta}<3.0$. The ECAL provides coverage in the pseudorapidity range $\abs{\eta}<1.5$ in the barrel (EB) region and $1.5<\abs{\eta}<3.0$ in the two endcap (EE) regions. The HCAL provides coverage for $\abs{\eta}<1.3$ in the barrel (HB) region and $1.3 <\abs{\eta}< 3.0$ in the two endcap (HE) regions. The hadron forward (HF) calorimeters ($3.0<\abs{\eta} < 5.2 $) complement the coverage provided by the barrel and endcap detectors. The zero-degree calorimeters (ZDCs) are two \v{C}erenkov calorimeters composed of alternating layers of tungsten and quartz fibers that cover the region $\abs{\eta}>8.3$. Both the HF and ZDC detectors are divided into two halves, one covering positive pseudorapidities, the other negative, and referred to as HF+ and ZDC+ (and HF- and ZDC-), respectively. Another calorimeter, CASTOR, also a \v{C}erenkov sampling calorimeter, consists of quartz and tungsten plates and is located only at negative pseudorapidities with coverage of $-6.6<\eta<-5.2$.
}
{\tolerance=40
A more detailed description of the CMS detector, together with the definition of the coordinate system used and the relevant kinematic variables, can be found in Ref.~\cite{Chatrchyan:2008zzk}.
}
\section{Data and Monte Carlo simulation}
\label{sec:Data}

{\tolerance=440
This analysis uses data from \Pp{}Pb collisions at $\sqrtsNN = 5.02\TeV$ collected with the CMS detector in February 2013. The beam energies are 4\TeV for the protons and 1.58\TeV per nucleon for the lead nuclei. The integrated luminosity is $\mathcal{L} = 7.4\mubinv$ for the \Pp{}Pb data set (protons circulating in the negative $z$ direction) and $\mathcal{L} = 9.6\mubinv$ for the Pb\Pp{} data set (protons circulating in the positive $z$ direction). Since the events are asymmetric in rapidity, the \Pp{}Pb and Pb\Pp~samples are merged after changing the sign of the rapidity in the Pb\Pp~sample.
}

{\tolerance=400
The\textsc{ starlight} (version 2.2.0) Monte Carlo (MC) event generator~\cite{Klein:2016yzr} is used to simulate exclusive $\Pgr^{0}$ photoproduction followed by the $\Pgr^{0}\to \Pgpp\Pgpm$ decay.
The\textsc{ starlight} generator models two-photon and photon-hadron interactions at ultrarelativistic energies. Two processes contribute to the exclusive $\Pgpp\Pgpm$ channel: resonant $\Pgr^{0}\to\Pgp^{+}\Pgp^{-}$ production, and nonresonant $\Pgpp\Pgpm$ production, including the interference term. Both processes are generated in order to calculate the signal acceptance and efficiency, and to extract the corrected signal yield.\textsc{ starlight} is also used to generate exclusive \Pgrb events. The \Pp{}Pb and Pb\Pp~samples are produced separately. The events are passed through a detailed \GEANTfour~\cite{Agostinelli:2002hh} simulation of the CMS detector in order to model the detector response, and are reconstructed with the same software used for the data.
}

\section{Event selection}
\label{sec:sele}

{\tolerance=400
Table~\ref{table:event_selection} presents the number of events after each selection requirement is applied.
Events were selected online~\cite{Khachatryan:2016bia} by requiring the simultaneous presence of the two beams at the interaction point, as measured by the beam monitor timing system, in conjunction with at least one track in the pixel tracker. Offline, events are discarded if they have an energy deposit in any of the HF towers above the noise threshold of 3\GeV. Events are also required to have exactly two tracks that pass the selection criteria defined in Ref.~\cite{Khachatryan:2010pw}, and to be associated with a single vertex located within 15\unit{cm} of the nominal interaction point along the beam direction. The pion mass is assigned to each track.
 In order to minimize the effect of the uncertainty in the low-\pt track
efficiency, one of the tracks should have a \pt larger than 0.4\GeV, and the other larger than 0.2\GeV. Both tracks are selected in the interval  $\abs{\eta} < 2.0$.
 The rapidity of the $\Pgpp\Pgpm$ system is required to be in the interval $\abs{y_{\Pgpp\Pgpm}}<2.0$.
To reject the photoproduction of $\Pgr^{0}$~mesons from $\Pgg\mathrm{Pb}$ interactions where the proton radiates a quasi-real photon, the \pt of the $\Pgpp\Pgpm$ system is required to be larger than $0.15$\GeV (as discussed in Section~\ref{sec:BG}).

A sizable background contribution comes from proton dissociative events, $\Pgg\Pp\to \Pgr^{0}\Pp^{*}$, where $\Pp^{*}$ indicates a low-mass hadronic state. In these events the scattered proton is excited and then dissociates.
The $\Pgr^{0}$ is measured in the central region, whereas the low-mass state usually escapes undetected. To suppress this contribution, events with activity above noise thresholds in the CASTOR, HE, HF, and ZDC detectors are rejected. The signal-to-noise ratio in ZDC$^{+}$ is better than in ZDC$^{-}$ because of differences in radiation damage to the two detectors. For this reason, the
ZDC energy thresholds shown in Table~\ref{table:event_selection} are asymmetric. CASTOR is used for only the pPb sample because of its location, as discussed in Section~\ref{sec:detector}.
The final selection requires the two tracks to have opposite charges.
A total of 20\,060 opposite-sign pair events and 1514 same-sign pair events are selected in this analysis.
}
\begin{table*}
\centering
\topcaption{\label{table:event_selection} Integrated luminosity and number of events after each of the selection requirements for the two data samples. The leading tower is the tower with the largest energy deposition in the calorimeter.}
\begin{tabular}{lrr}
Selection & Number of selected events &\\
        &       \Pp{}Pb & Pb\Pp \\
Integrated luminosity& 7.4 \mubinv& 9.6 \mubinv\\
\hline
Leading HF tower $<$ 3.0\GeV &52\,508 &66\,278\\
Exactly two tracks &    17\,771 & 21\,583\\
Track purity~\cite{Khachatryan:2010pw}   &   16\,085&20\,278\\
$\abs{\eta_\text{{track}}}<2.0$, & 12\,707&16\,037\\
$\pt^\text{{leading}} >0.4$\GeV, $\pt^{\text{subleading}}>0.2$\GeV & 12\,364&15\,572\\
$\abs{z_{\mathrm{vertex}}} < 15\cm$&   11\,924&15\,052\\
Leading HE tower $<$ 1.95\GeV&  11\,563&   14\,643\\
CASTOR energy $<$ 9\GeV&   9405&      \NA   \\
ZDC$^{+}$ energy $<$ 500\GeV&     \NA & 12\,475 \\
ZDC$^{-}$ energy $<$ 2000\GeV&  9099  & \NA\\[\cmsTabSkip]
Opposite-sign pairs &    8507&         11\,553\\
Same-sign pairs      & 592&            922\\
\hline
\end{tabular}
\end{table*}

\section{Background}
\label{sec:BG}

\begin{figure*}[ht!]
\includegraphics[width=0.8\textwidth]{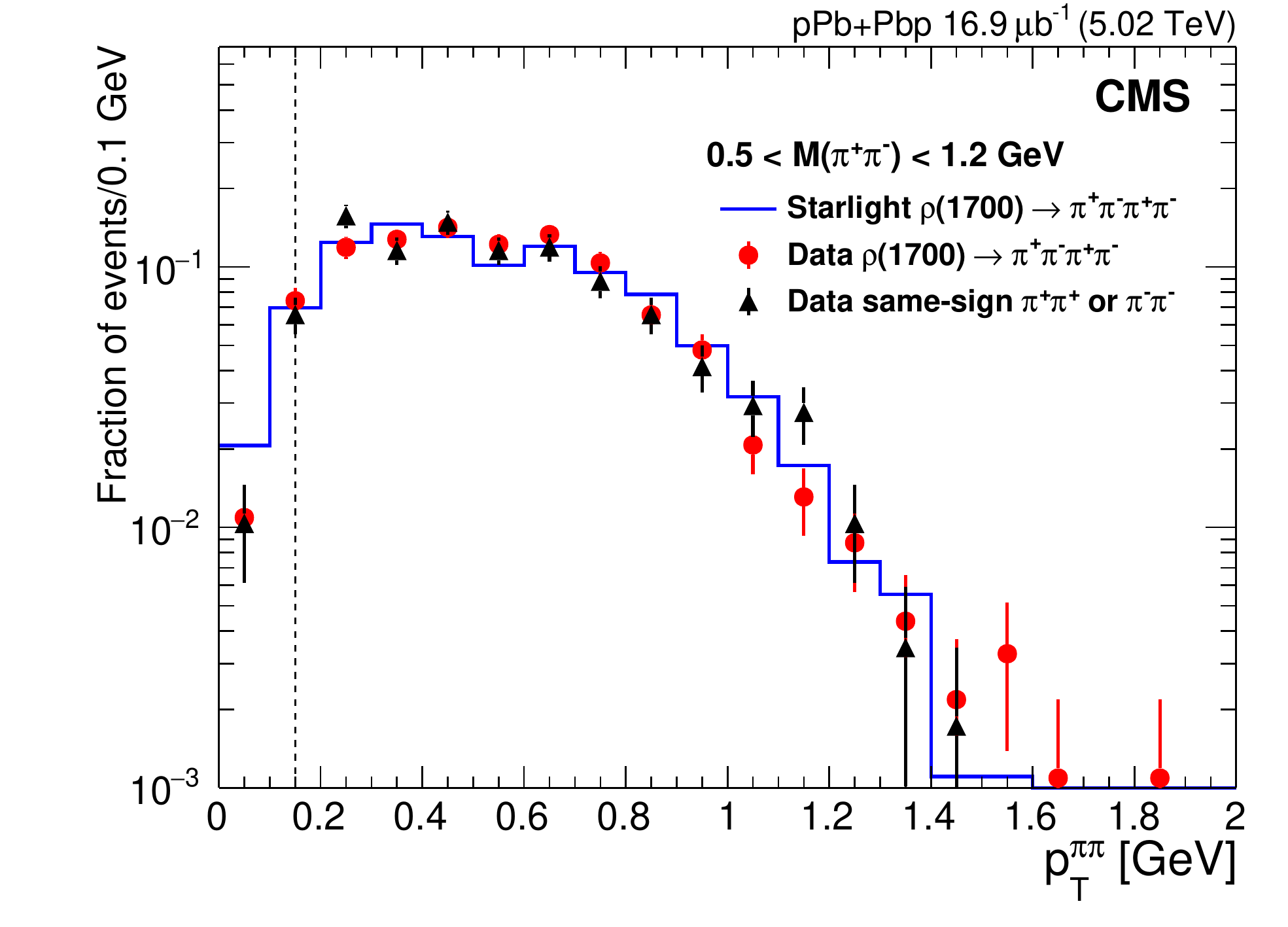}
 \caption[Template for the $\rho'$ background]{\label{figure:Rho_prime}
Comparison between the $\pt^{\Pgpp\Pgpm}$  distributions of the reconstructed $\Pgrb$ mesons in the data (full symbols) and the\textsc{ starlight} simulation (histogram) when only two oppositely charged pions are selected. The triangles correspond to same-sign two-track events (either $\Pgpp\Pgpp$ or $\Pgpm\Pgpm$) in the data; they mostly come from $\Pgrb$ decays with two undetected pions. The integrals of all three distributions are normalized to unity. Vertical bars correspond to the statistical uncertainties.
The region to the left of the dashed vertical line is not included in the analysis (see Section~\ref{sec:BG} for details).
}
\end{figure*}

{\tolerance=8000
The main background sources are listed below.

\begin{itemize}

\item \textit{Nonresonant $\Pgpp\Pgpm$ production}. This contributes mainly through an interference term. It is included when fitting
the invariant mass distribution, as discussed in Section~\ref{sec:minv_fit}.

\item \textit{Exclusive photoproduction of $\Pgo$ and $\Pgf$ mesons}. Contamination from the decay
$\Pgf\to\PKp\PKm$ is removed by assigning the kaon mass to the tracks and rejecting events with invariant mass values of the $\PKp\PKm$ system larger than 1.04\GeV. In addition, contamination is expected from the $\Pgo\to\Pgpp\Pgpm\Pgpz$ and $\Pgf\to\Pgpp\Pgpm\Pgpz$ decays when the photons from the \Pgpz~decay are undetected. Although the $\Pgpp\Pgpm$ invariant mass in these cases is mostly below the \Pgr~mass, the rate of $\Pgo$ and $\Pgf$ meson production increases with $\abs{t}$. As observed in this analysis and at HERA~\cite{Aaron:2009xp}, undetected photons lead to an overestimate of the \pt imbalance in the event, mimicking large $\abs{t}$ events. Since these processes cannot be modeled by\textsc{ starlight}, their contribution is estimated from the fits of the unfolded invariant mass distributions described in Section~\ref{sec:minv_fit}. The $\Pgo\to\Pgpp\Pgpm$ amplitude is small, but is clearly visible through its interference with the $\Pgr^{0}$, which produces the small kink in the invariant mass spectrum near 800\MeV. This contribution is included in the invariant mass fit, as discussed in Section~\ref{sec:minv_fit}.
\item \textit{Exclusive photoproduction of $\Pgrb$ mesons\footnotemark}. The $\Pgrb$ decays mostly into a $\Pgr^{0}$ meson and a pion pair, leading to final states with four charged pions, or with two charged pions and two neutral pions. The $\Pgrb\to \Pgpp\Pgpm\Pgpp\Pgpm$ decay may also result in opposite-sign events when only two opposite-sign pions are detected because of the limited rapidity coverage of the detector. Such events will appear to have a \pt imbalance, causing them to be incorrectly identified as large $\abs{t}~\Pgr^0$ events, thereby resulting in a distortion of the $\abs{t}$ distribution. To validate the use of\textsc{ starlight} for $\Pgrb$ photoproduction, exclusive $\Pgpp\Pgpm\Pgpp\Pgpm$ events are selected in the data. The data sample and the\textsc{ starlight} simulation for $\Pgrb$ exclusive photoproduction are studied by applying the same selection criteria as for the $\Pgr^{0}$, except that on the number of tracks. Figure~\ref{figure:Rho_prime} shows a comparison of the $\pt^{\Pgpp\Pgpm}$ distributions of the reconstructed $\Pgrb$ mesons in the four pion event samples obtained from the data and the\textsc{ starlight} simulation. All combinations of two oppositely charged pions are plotted in Fig.~\ref{figure:Rho_prime} if they have an invariant mass $0.5 < M_{\Pgpp\Pgpm} < 1.2\GeV$. In addition, the distribution of the same-sign events in the data is shown; they come mostly from $\Pgrb$ decays with two missing pions. Figure~\ref{figure:Rho_prime} shows that the data and the\textsc{ starlight} results are in agreement, lending confidence to the performance of this generator. These distributions provide a template for the $\pt^{\Pgpp\Pgpm}$ distribution of the $\Pgrb$ background used to estimate its contribution, as described in Section~\ref{sec:minv_fit}.
 \footnotetext{The data on the photoproduction of excited $\Pgrb$ states in the four-pion decay channel are currently limited. A resonance structure with a broad invariant mass distribution around 1600\MeV is reported in the literature. According to the Particle Data Group this resonance has two components: the $\Pgra$ and the $\Pgrb$~\cite{Tanabashi:2018oca}. The nature of these states is still under investigation. Recently, the STAR Collaboration reported a measurement of exclusive photoproduction of four charged pions~\cite{Abelev:2009aa}. Their data are consistent with the $\Pgrb$ assuming that the peak is dominated by spin states with $J^{PC} = 1^{--}$. In order to reproduce these data,\textsc{ starlight} assumes a single resonance with a mass of 1540\MeV and a width of 570\MeV~\cite{Klein:2016yzr}. In the present paper, this state is referred to as $\Pgrb$.}

\item \textit{Proton dissociative $\Pgr^{0}$~photoproduction}. This contribution is suppressed by rejecting events with activity in the CASTOR, HE, HF, and ZDC detectors. In order to determine the residual contribution, a sample of dissociative events is selected by requiring activity in at least one of the forward detectors (CASTOR, HF, or ZDC). This sample provides a template for the $\pt^{\Pgpp\Pgpm}$ distribution of the dissociative events, under the assumption that the $\pt^{\Pgpp\Pgpm}$ distribution is independent of the mass of the dissociative system (the more forward the detector, the smaller the masses to which it is sensitive). Finally, this template is used to estimate the remaining dissociative background contributions, as discussed in Section~\ref{sec:minv_fit}.

\item \textit{Double pomeron exchange processes and photoproduction processes from $\Pgg\mathrm{Pb}$ interactions}. Since the strong force has short range, only the nucleons on the surface of the nucleus may contribute to double pomeron exchange interactions; the corresponding cross section is therefore negligible~\cite{Schramm:1996aa}. For coherent processes in $\Pgg\mathrm{Pb}$ interactions, the size of the lead ion restricts the mean \pt of the VM to be about $60\MeV$, corresponding to a de Broglie wavelength of the order of the nucleus size. Taking into account the detector resolution, all coherent $\Pgr^{0}$ events have \pt less than $0.15\GeV$. Thus, events from $\Pgg\mathrm{Pb}$ interactions contribute to the lowest $\abs{t}$ region, which is not included in this analysis.
\end{itemize}
}

\section{Signal extraction}
\label{sec:minv_fit}

The extraction of the signal is carried out in two steps. First, the proton dissociative and the $\Pgrb$ contributions are estimated by performing a fit to the data as a function of $\pt^{\Pgpp\Pgpm}$.  This method relies on the
fact that exclusive $\Pgr^{0}$~events contribute mainly to the low-$\pt^{\Pgpp\Pgpm}$ region ($\pt^{\Pgpp\Pgpm}<0.7\GeV$), whereas nonexclusive events dominate the high-$\pt^{\Pgpp\Pgpm}$ region ($\pt^{\Pgpp\Pgpm}>1.2\GeV$), and the $\Pgrb$ contribution is mostly at intermediate $\pt^{\Pgpp\Pgpm}$ values ($0.7<\pt^{\Pgpp\Pgpm}<1.2\GeV$). This makes the identification of the proton dissociative and the $\Pgrb$ contributions robust.
Second, the yield of exclusive $\Pgr^{0}$~candidates is extracted by performing a fit to the unfolded invariant mass distribution.
Since the events from exclusive $\Pgrb$ production have a different invariant mass distribution from the signal events, they are subtracted before correcting the data for acceptance and efficiency. Conversely, the proton dissociative background has the same invariant mass and angular distributions as the signal, and its effect is corrected after unfolding by scaling the observed yields according to the fit performed in the first step.

{\tolerance=1200
To extract the normalizations of the proton dissociative and the $\Pgrb$ backgrounds, an unbinned maximum likelihood fit is performed to the data as a function of $\pt^{\Pgpp\Pgpm}$ in the rapidity interval $\abs{y_{\Pgpp\Pgpm}} < 2$.  The sum of the following distributions is fitted to the data at the reconstructed level: the signal distribution and the $\Pgpp\Pgpm$ continuum, as simulated by\textsc{ starlight}, the distribution of the proton dissociative background, which is extracted from the data control sample, and the $\Pgrb$ fitting template, which is simulated using \textsc{starlight}. The normalization of each of these components is determined from the fit. The signal $\pt^{\Pgpp\Pgpm}$ distribution generated by \textsc{starlight} is reweighted to describe the data using the theory-inspired expression $\re^{- b\abs{t}}$~\cite{Bauer:1977iq}. The initial $b$ value of \textsc{starlight} is $12\GeV^{-2}$ and the reweighted $b$ is $13.1^{+0.4}_{-0.3}\stat\GeV^{-2}$.
\par}

\begin{figure*}[ht!]
\includegraphics[width=1.0\textwidth]{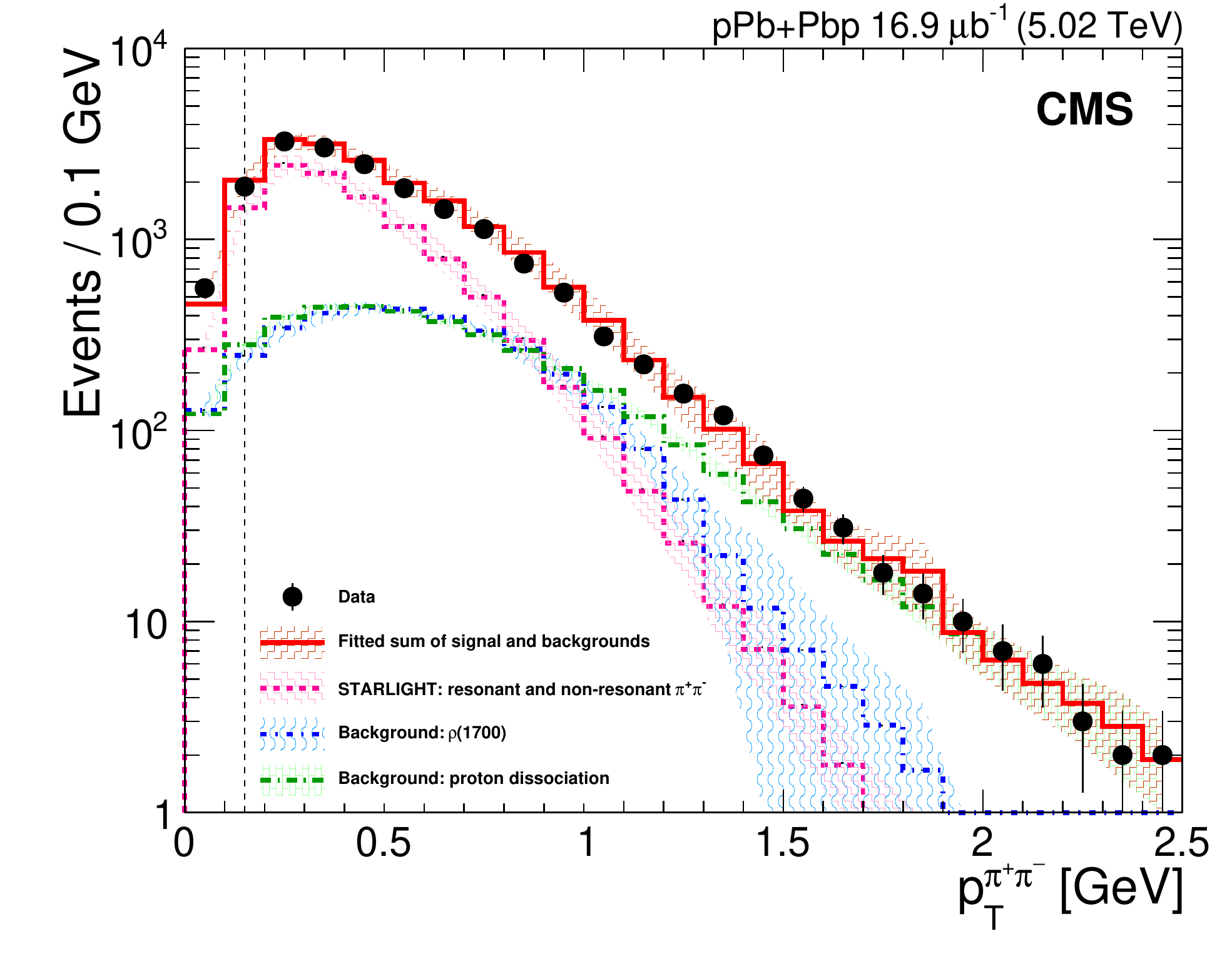}
\caption[Template fits to the reconstructed transverse momenta of $\Pgpp\Pgpm$ distribution]
{\label{figure:Roofit_Primepd_pt_allw}
The measured distribution of the reconstructed $\Pgpp\Pgpm$ transverse momentum (full circles) together with the fitted sum of signal and backgrounds described in the text (red solid histogram). The\textsc{ starlight} direct $\Pgpp\Pgpm$ contribution (pink dotted histogram), the $\Pgrb$ background (blue dotted-short-dashed histogram), and the proton-dissociative contribution (green dotted-long-dashed histogram) are also shown. The shaded areas represent the systematic uncertainties.
The region to the left of the dashed vertical line is not included in the analysis (see Section~\ref{sec:BG} for details).
}
\end{figure*}

{\tolerance=300
The result of the fit of the $\pt^{\Pgpp\Pgpm}$ distributions is shown in Fig.~\ref{figure:Roofit_Primepd_pt_allw}, including the systematic uncertainties associated with the fitting procedure that are discussed in Section~\ref{sec:xsec2}. The resulting residual proton-dissociative and $\Pgrb$ contributions, over the whole rapidity interval, are $18\pm2\%$ \stat and $20\pm2\%\stat$, respectively. Similar fractions of proton dissociative and $\Pgrb$ contributions are obtained in the four rapidity intervals used in the differential cross section measurement as a function of rapidity. This is consistent with the small energy dependence of these processes in the energy range of this analysis. As seen in Fig.~\ref{figure:Roofit_Primepd_pt_allw}, the signal and both background contributions are of the same order of magnitude around $\pt^{\Pgpp\Pgpm}=1\GeV$,  corresponding to a signal-to-background ratio of about 30\%. For this reason, only the region $\abs{t} < 1$ $\GeVns^{2}$ is used in this measurement.
}

The $\Pgrb$ background is subtracted in bins of invariant mass using the normalization obtained from the $\pt^{\Pgpp\Pgpm}$ fitting templates. The invariant mass distribution is then unfolded using the iterative D'Agostini method~\cite{DAGOSTINI1995487}, which is regularized by four iterations. In particular, the Bayesian iterative unfolding technique is used, as implemented in the\textsc{ roounfold} package~\cite{Adye:2011gm}. This procedure leads to corrections for experimental effects including possible data migration between bins.
The response matrix is obtained from\textsc{ starlight}. The average of the combined acceptance and efficiency is 0.13 and is almost independent of both \pt and $\eta$, whereas it is sensitive to the invariant mass.

The  invariant mass shape  of  the $\Pgr^{0}$ in photoproduction  deviates  from  that of a pure Breit--Wigner resonance~\cite{Mcclellan:1972cz}.
Several parameterizations of the shape exist. One of the most often used is the S\"{o}ding formula~\cite{Soding:1965nh}, where a continuum amplitude $B$ is added to a Breit--Wigner distribution. Following the recent results by the STAR Collaboration~\cite{Adamczyk:2017vfu} and the earlier ones by the DESY-MIT Collaboration~\cite{Alvensleben:1971hz}, a further relativistic Breit-Wigner component is added to account for $\Pgo$ photoproduction, followed by the decay $\Pgo\to\Pgpp\Pgpm$. This leads to the following fitting function:
\begin{linenomath}
\ifthenelse{\boolean{cms@external}}
{ 
\begin{multline*}
\frac{\rd N_{\Pgpp\Pgpm}}{\rd M_{\Pgpp\Pgpm}} =
\Biggl|
A\frac{\sqrt{M_{\Pgpp\Pgpm}M_{\Pgr}\Gamma_{\Pgr}}}{M_{\Pgpp\Pgpm}^{2} - M_{\Pgr^{0}}^{2} + iM_{\Pgr^{0}}\Gamma_{\Pgr}}\\
+ B
+ C\re^{i\phi_\omega}\frac{\sqrt{M_{\Pgpp\Pgpm}M_{\Pgo}\Gamma_{\Pgo\to\Pgp\Pgp}}}{M_{\Pgpp\Pgpm}^{2} - M_{\Pgo}^{2} + iM_{\Pgo^{0}}\Gamma_{\Pgo}}
\Biggr|^{2}.
\end{multline*}
} 
{ 
\begin{equation*}
\label{eq:Soding}
\frac{\rd N_{\Pgpp\Pgpm}}{\rd M_{\Pgpp\Pgpm}} =
\left| A\frac{\sqrt{M_{\Pgpp\Pgpm}M_{\Pgr}\Gamma_{\Pgr}}}{M_{\Pgpp\Pgpm}^{2} - M_{\Pgr^{0}}^{2} + iM_{\Pgr^{0}}\Gamma_{\Pgr}}
+ B
+ C\re^{i\phi_\omega}\frac{\sqrt{M_{\Pgpp\Pgpm}M_{\Pgo}\Gamma_{\Pgo\to\Pgp\Pgp}}}{M_{\Pgpp\Pgpm}^{2} - M_{\Pgo}^{2} + iM_{\Pgo^{0}}\Gamma_{\Pgo}} \right|^{2}.
\end{equation*}
} 
\end{linenomath}
Here $A$ is the amplitude of the $\Pgr^{0}$ Breit--Wigner function, $B$ is the amplitude of the direct nonresonant
$\Pgpp\Pgpm$~production, $C$ is the amplitude of the \Pgo contribution, and the mass-dependent widths are given by
\begin{equation*}
\Gamma_{\Pgr} = \Gamma_{0}\frac{M_{\Pgr^{0}}}{M_{\Pgpp\Pgpm}} \left[ \frac{M_{\Pgpp\Pgpm}^{2} - 4m_{\Pgppm}^{2}}{M_{\Pgr^{0}}^{2} - 4m_{\Pgppm}^{2}} \right] ^{\frac{3}{2}},
\end{equation*}
and
\begin{equation*}
\Gamma_{\Pgo} = \Gamma_{0}\frac{M_{\Pgo}}{M_{\Pgpp\Pgpm}} \left[ \frac{M_{\Pgpp\Pgpm}^{2} - 9m_{\Pgppm}^{2}}{M_{\Pgo}^{2} - 9m_{\Pgppm}^{2}} \right] ^{\frac{3}{2}},
\end{equation*}
where $\Gamma_{0}$ is the pole width for each meson and $m_{\Pgppm}$ is the charged pion mass. Since the branching fraction ($\mathcal{B}$) for $\Pgo\to\Pgpp\Pgpm$ is small, only the first order term in the $\Pgo-\Pgr^{0}$ mass mixing theory is considered~\cite{Alvensleben:1971hz}, leading to
\begin{linenomath}
\ifthenelse{\boolean{cms@external}}
{ 
\begin{multline*}
\Gamma_{\Pgo\to\Pgp\Pgp} = \\
\mathcal{B}(\Pgo\to\Pgp\Pgp)\Gamma_{0}\frac{M_{\Pgo}}{M_{\Pgpp\Pgpm}}
\left[ \frac{M_{\Pgpp\Pgpm}^{2} - 4m_{\Pgppm}^{2}}{M_{\Pgo}^{2} - 4m_{\Pgppm}^{2}} \right] ^{\frac{3}{2}},
\end{multline*}
} 
{ 
\begin{equation*}
\Gamma_{\Pgo\to\Pgp\Pgp} = \mathcal{B}(\Pgo\to\Pgp\Pgp)\Gamma_{0}\frac{M_{\Pgo}}{M_{\Pgpp\Pgpm}} \left[ \frac{M_{\Pgpp\Pgpm}^{2} - 4m_{\Pgppm}^{2}}{M_{\Pgo}^{2} - 4m_{\Pgppm}^{2}} \right] ^{\frac{3}{2}},
\end{equation*}
}
\end{linenomath}
with $\mathcal{B}(\Pgo\to\Pgp\Pgp)=0.0153^{+0.0011}_{-0.0013}$~\cite{Tanabashi:2018oca}. The H1 and ZEUS measurements did not include the $\Pgo-\Pgr^{0}$ interference
component, although the ZEUS data seem to indicate its effect in the mass spectrum near 800\MeV~\cite{Breitweg:1997ed}.

\begin{figure*}[ht!]
\centering
  \includegraphics[width=.8\textwidth]{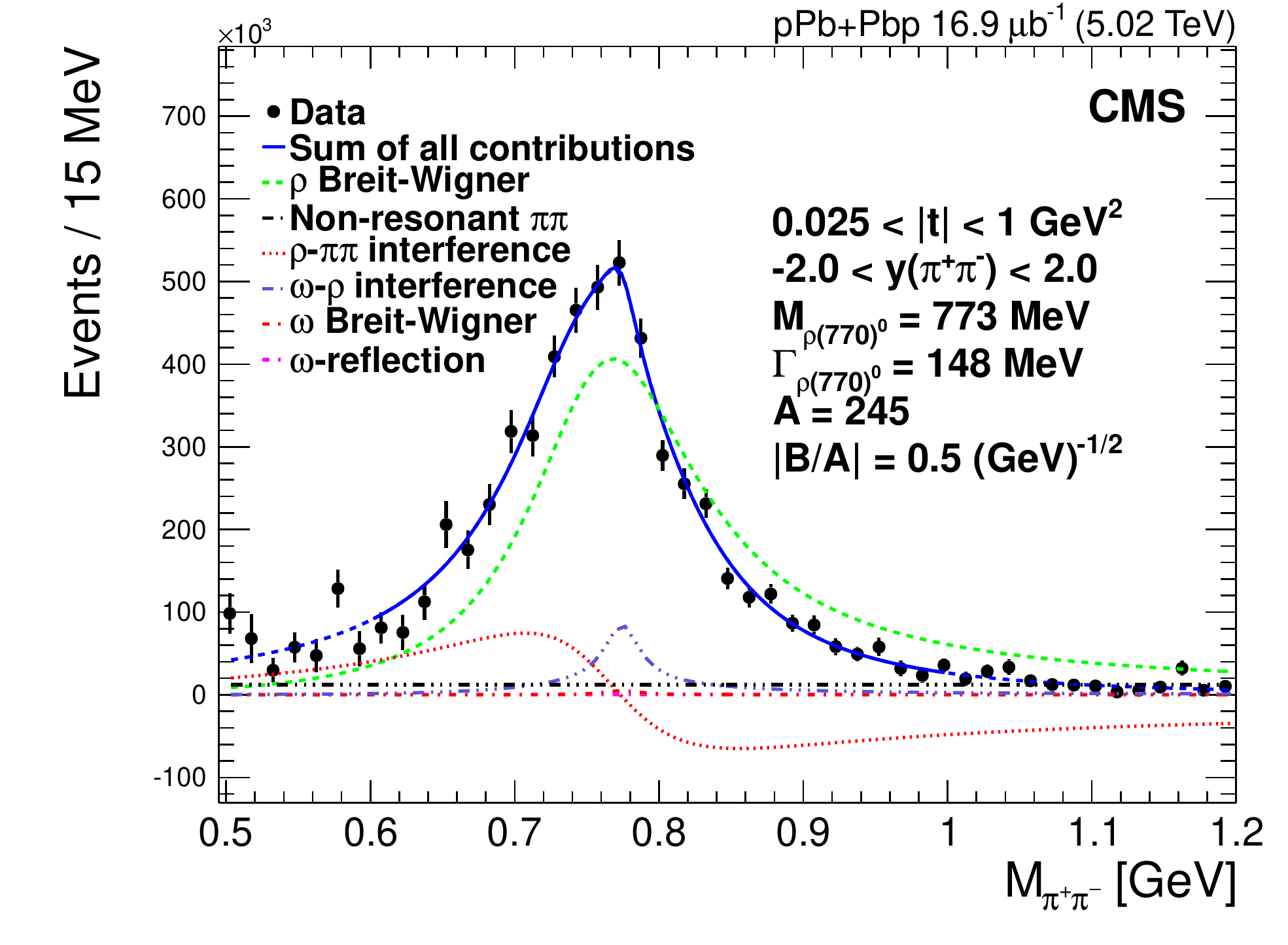}
 \caption[Invariant mass $M_{inv} (\Pgpp\Pgpm)$ distribution for $\Pgr^{0}$~in the rapidity
interval $\abs{y(\Pgpp\Pgpm)} < 2.0$ fitted with the S\"{o}ding model~\ref{eq:Soding}]{\label{figure:mass_fit_all_all}
Unfolded $\Pgpp\Pgpm$ invariant mass distribution in the pion pair rapidity interval $\abs{y_{\Pgpp\Pgpm}} < 2.0$
(full circles) fitted with the modified S\"{o}ding model. The results of the fit are also given (see text for details). The green dashed line indicates resonant $\Pgr^{0}$~production, the red dotted line the interference term,
the black dash-dotted line the non-resonant contribution, the dark blue dashed line the interference between $\Pgr^{0}$ and $\Pgo$, and
the blue solid line represents the sum of all these contributions.}
\end{figure*}

\begin{figure*}[ht!]
\centering
  \includegraphics[width=.95\textwidth]{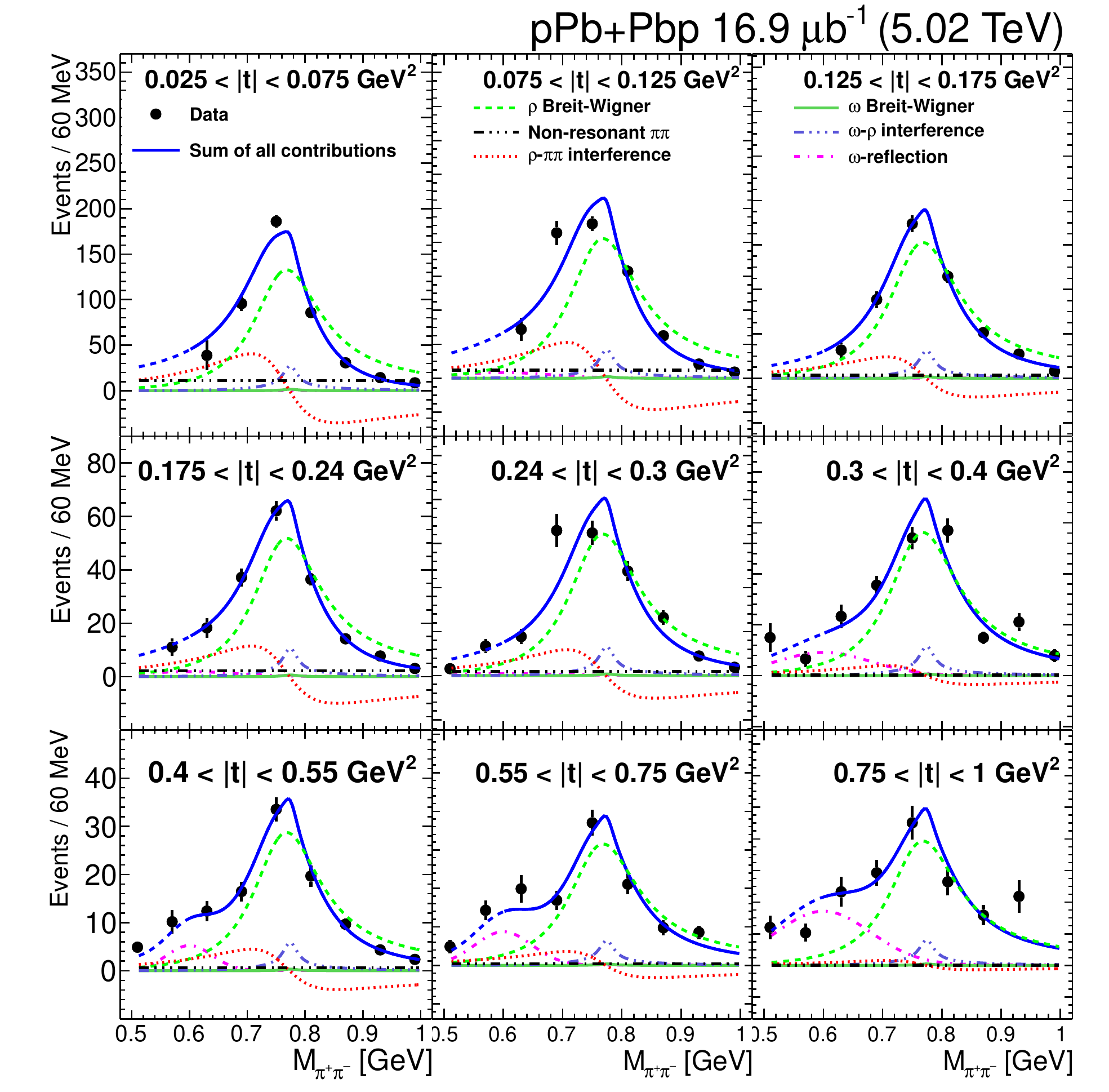}
 \caption[Invariant mass $M_{inv} (\Pgpp\Pgpm)$ distribution for $\Pgr^{0}$~in the rapidity
interval $\abs{y(\Pgpp\Pgpm)} < 2.0$ fitted with the S\"{o}ding model~\ref{eq:Soding}]{\label{figure:mass_fit_all}
Unfolded $\Pgpp\Pgpm$ invariant mass distributions in the pion pair rapidity interval $\abs{y_{\Pgpp\Pgpm}} < 2.0$
(full circles) fitted with the S\"{o}ding model in different $\abs{t}$ bins. The green dashed lines indicate resonant $\Pgr^{0}$~production, the red dotted lines the interference term,
the magenta dash-dotted lines correspond to the background from $\Pgo\to\Pgpz\Pgpp\Pgpm$, the black dash-dotted lines to the nonresonant contribution, the dark blue dashed line to the interference between $\Pgr^{0}$ and $\Pgo$, and
the blue solid lines represent the sum of all these contributions.}
\end{figure*}

Figure~\ref{figure:mass_fit_all_all} shows the fit of the unfolded distribution with the modifed S\"{o}ding model. 
A least squares fit is performed for the interval $0.6 < M_{\Pgpp\Pgpm} < 1.1\GeV$, with the quantities $M_{\Pgr^{0}}$, $M_{\Pgo}$, $\Gamma_{\Pgr^{0}}$, $\Gamma_{\Pgo}$, $A$, $B$, $C$, and $\phi_\Pgo$ treated as free parameters. This model includes the interference between resonant $\Pgr^{0}$~and direct $\Pgp^{+}\Pgp^{-}$~production, as well as between $\Pgr^{0}$ and $\Pgo$ production.
To correct for the \Pgo~reflection in the $\Pgpp\Pgpm$ mass spectrum, a Gaussian function peaking around 500\MeV~\cite{Aid:1996bs} is added as a further component of the invariant mass fit. This is only visible at high $\abs{t}$ values, as shown in Fig.~\ref{figure:mass_fit_all}.
The fit yields $M_{\Pgr^{0}} = 773\pm1$\stat\MeV and $\Gamma_{\Pgr^{0}} = 148\pm3\stat\MeV$, and $M_{\Pgo} = 776\pm2$\stat\MeV, consistent with the world average values~\cite{Tanabashi:2018oca}. The fitted value of the \Pgo width, $\Gamma_{\Pgo} = 30\pm5$\stat\MeV, is instead larger than the world average because of the detector resolution. 

The $\abs{B/A}$ and $C/A$ fractions are also determined; they measure the ratios of the nonresonant and \Pgo contributions to the resonant $\Pgr^{0}$ production, respectively. Since the ZEUS Collaboration found that $\abs{B/A}$ decreases as $\abs{t}$ increases, the fit is repeated for $\abs{t}<0.5$ $\GeVns^{2}$ resulting in $0.50 \pm 0.06$\stat$\GeVns^{-1/2}$. For this kinematic region H1 measured $\abs{B/A}=0.57 \pm 0.09$\stat$\GeVns^{-1/2}$ and ZEUS $\abs{B/A}=0.70 \pm 0.04$\stat$\GeVns^{-1/2}$. If the fit is repeated without the $\Pgo-\Pgr^{0}$ interference component, the result for $\abs{B/A}$ changes by less than its statistical uncertainty.
The measured ratio of the \Pgo to $\Pgr^{0}$ amplitudes is $C/A=0.40\pm0.06$\stat, consistent with the prediction of \textsc{starlight}, $C/A=0.32$, and the measurements of the STAR~\cite{Adamczyk:2017vfu} and the \textsc{DESY-MIT}~\cite{Alvensleben:1971hz} experiments, which report $C/A=0.36\pm0.03$\stat and $C/A=0.36\pm0.04$\stat, respectively. The present fit gives a nonzero \Pgo phase angle, $\phi_\Pgo=1.8\pm0.3$\stat, also in agreement with the previous measurements~\cite{Adamczyk:2017vfu,Alvensleben:1971hz}.

 Additionally, the fit is performed in $\abs{t}$ and $y$ bins as shown in Fig.~\ref{figure:mass_fit_all}. To ensure fit stability, the $M_{\Pgr^{0}}$, $M_{\Pgo}$, $\Gamma_{\Pgr^{0}}$, $\Gamma_{\Pgo}$, $\phi_\Pgo$ and $\abs{C/A}$ parameters are fixed to the values obtained for the full rapidity interval.
 The $\Pgo\to\Pgpp\Pgpm\Pgpz$ contribution increases with $\abs{t}$, as reported by the H1 Collaboration~\cite{Aid:1996bs} and as seen in Fig.~\ref{figure:mass_fit_all}.  The $\abs{B/A}$ ratio is found to be independent of $W_{\Pgg\Pp}$ and decreases with $\abs{t}$, in agreement with results reported by ZEUS~\cite{Breitweg:1997ed}.

\section{Systematic uncertainties}
\label{sec:xsec2}

The following sources of systematic uncertainty are considered.

 \textit{ Integrated luminosity determination}: The uncertainty in the integrated luminosity is
4\% for both the pPb and Pbp samples~\cite{CMS:2014uoa}.

 \textit{ Track reconstruction}: The contribution of the tracking efficiency to the systematic uncertainty is studied with the method described in Ref.~\cite{Chatrchyan:2014fea},
where the ratio of yields of neutral charm mesons decaying to two-body and four-body final states is compared with
data and simulation for pion momenta above 300\MeV. The accuracy of the detector simulation to reproduce the single-pion
tracking efficiency is 3.9\%. For the present measurement, this yields a 7.8\% uncertainty.

 \textit{ Unfolding}: The uncertainty associated with the unfolding procedure is determined by modifying the number of iterations used for the Bayesian unfolding from the nominal value of 4 to 3 and 5. The resulting uncertainty is smaller than that found when changing the model for building the response matrix. The latter is estimated by comparing two different\textsc{ starlight} samples: resonant $\Pgr^{0}$~meson production, and combined resonant and nonresonant $\Pgpp\Pgpm$ production. The resulting effect on the integrated cross section is 3\%.

 \textit{ Uncertainty in the photon flux:} The uncertainty in the photon flux is 9\% for the high-$W_{\Pgg\Pp}$ data point and 2\% at low $W_{\Pgg\Pp}$, as discussed in Ref.~\cite{TheALICE:2014dwa}.
The flux is computed in impact parameter space, convolved with the probability of no hadronic interactions. The radius of the lead nucleus is varied by the nuclear skin thickness ($\pm 0.5\unit{fm}$). In addition, in the calculation of the photon flux, the $\Pgr^{0}$~pole mass in Eq.~(\ref{eq:flux}) is replaced by the reconstructed $\Pgpp\Pgpm$ mass on an event-by-event basis. The effect of this variation is negligible.

 \textit{ Calorimeter exclusivity}: The uncertainty related to the exclusivity requirements is evaluated by varying the calorimeter energy thresholds.
Increasing (or decreasing) the energy scale of the HF calorimeter towers by 5\% results in a 1.0\% variation of the
exclusive $\Pgpp\Pgpm$ yields. The CASTOR energy scale is varied by $17\%$~\cite{Sirunyan:2017nsj}, resulting in a difference of 1\% in the extracted $\Pgr^{0}$~yield. The variations of the energy thresholds for HE and ZDC within their respective energy scale uncertainties have a negligible effect.

 \textit{ Background estimation}: The uncertainty in the $\Pgrb$ subtraction is evaluated by varying the normalization of the $\Pgrb$ contribution by 20\% with respect to that obtained from the fit shown in Fig.~\ref{figure:Roofit_Primepd_pt_allw}.
As mentioned in Section 5, the proton dissociative background template is obtained by requiring a signal in at least one of the forward detectors:  HF, CASTOR, or ZDC. To calculate the systematic uncertainty related to the estimation of this background, the analysis is repeated five times and each time alternative combinations of forward detectors are used to obtain the proton dissociative template. The following variations are studied: \textit{i)} HF alone; \textit{ii)} CASTOR alone; \textit{iii)} ZDC alone; \textit{iv)} HF or CASTOR; \textit{v)} HF or ZDC. For each of these combinations the proton dissociative contributions are obtained in each $\abs{t}$ and rapidity bin.  The maximum deviations from the nominal results are taken as conservative estimates of the systematic uncertainty.
The resulting effect on the integrated exclusive $\Pgr^{0}$~photoproduction cross section is smaller than 10\%.

 \textit{ Model dependence}: In order to assess the uncertainty due to the model used to fit the invariant mass distribution,
the Ross--Stodolsky model~\cite{Ross:1965qa} is used instead of the S\"{o}ding model. The resulting cross section changes by up to
8\%, depending on the rapidity and $\abs{t}$ interval studied.
Another contribution to the model dependence uncertainty comes from the reweighting procedure of the\textsc{ starlight} MC described in Section~\ref{sec:minv_fit}. This uncertainty is evaluated by varying the reweighting parameter $b$ within its uncertainty; it is found to increase as a function of $\abs{t}$, and reaches 32\% for the highest $\abs{t}$ bin. The second contribution turns out to be dominant for all the rapidity and $\abs{t}$ intervals studied.
The uncertainty in the extrapolation to the region $\abs{t} < 0.025\GeV^{2}$  is model dependent. We estimated this uncertainty by studying different fitting functions to the differential cross section measurements. In particular, we studied a dipole form~\cite{Klein:2016yzr}, a pure exponential $\re^{-bt}$, and a modified exponential $\re^{-bt+ct^2}$. The difference between the two most extreme extrapolated values is used as an estimate of the model dependence uncertainty.

The values of the systematic uncertainties for all $y_{\Pgpp\Pgpm}$ and $\abs{t}$ intervals are summarized in Table~\ref{tab:syst}.
The systematic uncertainties are added in quadrature for the integrated photoproduction cross section. For the differential cross section results, the systematic uncertainties in Table~\ref{tab:syst} are treated as correlated between bins.
\begin{table*}
\centering
\topcaption{Summary of the systematic uncertainties in the $\Pgr^{0}$ photoproduction cross section. The numbers are given in percent. The total uncertainty is calculated by adding the individual uncertainties in quadrature.}
\label{tab:syst}
\begin{tabular}{lrrrrr}
\hline
$y_{\Pgpp\Pgpm}$ interval 	 & $(-2.0, 2.0)$& $(-2.0, -1.2)$& $(-1.2, 0.0)$& $(0.0, 1.2)$& $(1.2, 2.0)$\\
\hline
Integrated luminosity &4.0&4.0&4.0&4.0&4.0\\
Track reconstruction &7.8&7.8&7.8&7.8&7.8\\
Unfolding &3.0&3.0&3.0&3.0&3.0\\
Photon flux calculation &5.0&2.0&4.0&6.0&9.0\\
Calorimeter exclusivity &1.4&1.4&1.4&1.4&1.4\\[\cmsTabSkip]

proton dissociation&&&&&\\
$\abs{t} [\GeVns^{2}]$&&&&&\\
0.025--1.000 & 2.3 & 1.6 & 2.3 & 3.6 & 3.9\\
0.025--0.075 & 2.3 & 1.6 & 2.3 & 3.6 & 3.9\\
0.075--0.120 & 1.9 & 1.5 & 1.8 & 2.9 & 3.2\\
0.12--0.17 & 2.3 & 1.7 & 2.1 & 3.3 & 3.7\\
0.17--0.24 & 3.0 & 2.2 & 2.7 & 4.0 & 4.9\\
0.24--0.30 & 3.9 & 2.5 & 3.4 & 5.5 & 6.8\\
0.3--0.4 & 5.2 & 3.7 & 4.6 & 7.1 & 9.2\\
0.40--0.55 & 7.1 & 5.8 & 6.5 & 9.8 & 13.0\\
0.55--0.75 & 10.0 & 9.7 & 9.0 & 14.0 & 19.0\\
0.75--1.00 & 14.0 & 19.0 & 11.0 & 22.0 & 28.0\\[\cmsTabSkip]

$\Pgrb$ background&&&&&\\
$\abs{t} [\GeVns^{2}]$&&&&&\\
0.025--1.000 & 4.3 & 13.0 & 1.9 & 7.1 & 2.0\\
0.025--0.075 & 4.3 & 13.0 & 1.9 & 7.1 & 2.0\\
0.075--0.120 & 4.8 & 1.3 & 2.0 & 2.0 & 7.4\\
0.12--0.17 & 5.6 & 2.9 & 2.7 & 4.5 & 5.6\\
0.17--0.24 & 5.8 & 3.6 & 5.9 & 3.2 & 4.9\\
0.24--0.30 & 3.8 & 4.4 & 6.6 & 5.8 & 16.0\\
0.3--0.4 & 6.5 & 14.0 & 7.3 & 11.0 & 17.0\\
0.40--0.55 & 9.1 & 19.0 & 21.0 & 9.7 & 14.0\\
0.55--0.75 & 35.0 & 37.0 & 13.0 & 20.0 & 55.0\\
0.75--1.00 & 46.0 & 56.0 & 19.0 & 39.0 & 32.0\\[\cmsTabSkip]

Model dependence&&&&&\\
$\abs{t} [\GeVns^{2}]$&&&&&\\
0.025--1.000 & 5.1 & 5.1 & 5.1 & 5.1 & 5.1\\
0.025--0.075 & 5.1 & 5.1 & 5.1 & 5.1 & 5.1\\
0.075--0.120 & 5.5 & 5.5 & 5.5 & 5.5 & 5.5\\
0.12--0.17 & 6.6 & 6.6 & 6.6 & 6.6 & 6.5\\
0.17--0.24 & 8.6 & 8.6 & 8.6 & 8.6 & 8.6\\
0.24--0.30 & 12.0 & 12.0 & 12.0 & 12.0 & 11.0\\
0.3--0.4 & 15.0 & 16.0 & 16.0 & 15.0 & 15.0\\
0.40--0.55 & 20.0 & 20.0 & 20.0 & 20.0 & 20.0\\
0.55--0.75 & 26.0 & 26.0 & 26.0 & 26.0 & 25.0\\
0.75--1.00 & 32.0 & 32.0 & 32.0 & 32.0 & 32.0\\
\hline
\end{tabular}
\end{table*}

\section{Results}
\label{sec:xsec}

{\tolerance=400
The differential cross section for exclusive photoproduction of $\Pgr^{0}$~mesons is given by
\begin{equation*}
\frac{\rd\sigma}{\rd y} = \frac{N^{\text{exc}}_{\Pgr^{0}}}{\mathcal{B}(\Pgr^{0}\to\Pgp^{+}\Pgp^{-})  L  \Delta y},
\end{equation*}
\noindent
where $N^{\text{exc}}_{\Pgr^{0}}$ is the corrected number of exclusive $\Pgr^{0}$ events obtained from the fits described in Section~\ref{sec:minv_fit} by
integrating the resonant component in the interval $0.28 < M_{\Pgr^{0}}<1.50\GeV$ ($2M_{\Pgppm} < M_{\Pgr^{0}} < M_{\Pgr^{0}} + 5\Gamma_{\Pgr^{0}}$); $\mathcal{B}$ is the branching fraction, which equals about $0.99$ for the $\Pgr^{0}\to\Pgp^{+}\Pgp^{-}$ decay~\cite{Tanabashi:2018oca}, $\Delta y$ is the rapidity interval, and $L$ is the integrated luminosity of the data sample. The cross section
$\rd \sigma/\rd y(\Pp\mathrm{Pb}\to\Pp\mathrm{Pb}\Pgr^{0})$ is related to the photon-proton cross section,
$\sigma(\Pgg\Pp\to \Pgr^{0}\Pp) \equiv \sigma(W_{\Pgg\Pp})$, through the photon flux, $\rd n/\rd k$:
\begin{equation*}
\frac{\rd \sigma}{\rd y}(\Pp\mathrm{Pb}\to\Pp\mathrm{Pb}\Pgr^{0}) = k \frac{\rd n}{\rd k}
\sigma(\Pgg\Pp\to \Pgr^{0}\Pp).
\end{equation*}
\noindent

Here, $k$ is the photon energy, which is determined from the $\Pgr^{0}$~mass and rapidity, according to the formula
\begin{equation}
\label{eq:flux}
 k = (1/2) M_{\Pgr^{0}} \exp{(-y_{\Pgr^{0}})}.
\end{equation}
 The average photon flux and the average centre-of-mass energy ($\langle W_{\Pgg\Pp} \rangle$) values in each rapidity interval are calculated using \textsc{ starlight}.
}
\begin{table*}
\centering
\topcaption{Differential cross section for exclusive $\Pgr^{0}$~photoproduction, $\sigma(\Pgg\Pp\to \Pgr^{0}\Pp)$, with statistical and
systematic uncertainties, for $\abs{t}<0.5$ $\GeVns^{2}$. The differential cross section $\rd\sigma/\rd\abs{t}$ is also shown, along with the rapidity range, the average value of $W_{\Pgg\Pp}$, $\langle W_{\Pgg\Pp}\rangle$, and $ k \frac{\rd n}{\rd k}$.}
\label{tab:xs}
\resizebox{\textwidth}{!}{\begin{tabular}{lrrrrr}
\hline
$y$ range&(-2.0, 2.0)& (-2.0, -1.2)& (-1.2, 0.0)& (0.0, 1.2)& (1.2, 2.0) \\
$W_{\Pgg\Pp}$ range [\GeVns]&(29, 213) &(29, 43) & (43, 78) & (78, 143) & (143, 213)\\
$\langle W_{\Pgg\Pp}\rangle$ [\GeVns]&92.6 & 35.6 & 59.2 & 108.0 & 176.0\\
$ k \frac{\rd n}{\rd k}$&136.0 & 186.0 & 155.0 & 117.0 & 86.2\\[\cmsTabSkip]

$\rd\sigma/\rd y$ [$\mu$b]&11.0 & 9.1 & 9.9 & 12.4 & 12.9\\
Stat. unc. [$\mu$b]&1.4 & 1.5 & 1.6 & 2.4 & 2.6\\
Syst. unc. [$\mu$b]&1.0 & 0.8 & 0.9 & 1.1 & 1.3\\[\cmsTabSkip]

$\abs{t} $[\GeVns$^{2}$]&$\rd\sigma/\rd\abs{t}$ [$\mu$b/\GeVns$^{2}$] &$\rd\sigma/\rd\abs{t}$ [$\mu$b/\GeVns$^{2}$]&$\rd\sigma/\rd\abs{t}$ [$\mu$b/\GeVns$^{2}$]&$\rd\sigma/\rd\abs{t}$ [$\mu$b/\GeVns$^{2}$]&$\rd\sigma/\rd\abs{t}$ [$\mu$b/\GeVns$^{2}$]\\[\cmsTabSkip]

0.025--0.075&56.0$\pm$2.2$\pm$6.4&47.0$\pm$4.5$\pm$4.9&50.0$\pm$4.1$\pm$5.5&57.7$\pm$6.1$\pm$6.9&74.5$\pm$7.9$\pm$10.2\\
0.075--0.125&33.6$\pm$1.0$\pm$3.9&26.0$\pm$2.3$\pm$2.8&30.2$\pm$1.9$\pm$3.4&39.1$\pm$3.2$\pm$4.7&39.3$\pm$3.4$\pm$5.5\\
0.125--0.175&24.4$\pm$0.8$\pm$3.0&22.1$\pm$2.1$\pm$2.6&18.8$\pm$1.2$\pm$2.3&24.3$\pm$2.2$\pm$3.1&26.6$\pm$2.3$\pm$3.9\\
0.175--0.240&15.5$\pm$0.7$\pm$2.1&10.9$\pm$1.3$\pm$1.4&14.6$\pm$1.2$\pm$2.0&16.5$\pm$1.9$\pm$2.4&14.1$\pm$1.7$\pm$2.2\\
0.24--0.30&10.2$\pm$0.6$\pm$1.6&6.7$\pm$0.8$\pm$1.0&9.7$\pm$0.9$\pm$1.5& 11.8 $\pm$1.9$\pm$1.9&8.1$\pm$1.1$\pm$1.4\\
0.3--0.4&5.2$\pm$0.4$\pm$1.0&5.0$\pm$0.9$\pm$0.9&4.0$\pm$0.5$\pm$0.8&6.6$\pm$1.5$\pm$1.3&3.3$\pm$0.6$\pm$0.7\\
0.40--0.55&3.5$\pm$0.4$\pm$0.8&2.2$\pm$0.6$\pm$0.5&3.4$\pm$0.6$\pm$0.8&3.0$\pm$1.0$\pm$0.7&1.9$\pm$0.5$\pm$0.5\\
0.55--0.75&1.4$\pm$0.3$\pm$0.5&0.94$\pm$0.44$\pm$0.37&1.5$\pm$0.3$\pm$0.6&1.2$\pm$0.6$\pm$0.5&1.0$\pm$0.3$\pm$0.4\\
0.75--1.00&0.52$\pm$0.14$\pm$0.27&0.37$\pm$0.28$\pm$0.19&0.50$\pm$0.12$\pm$0.26&0.60$\pm$0.47$\pm$0.31&0.38$\pm$0.22$\pm$0.20\\
\hline
\end{tabular}}
\end{table*}
\begin{figure*}[ht!]
\quad
\includegraphics[width=.45\textwidth]{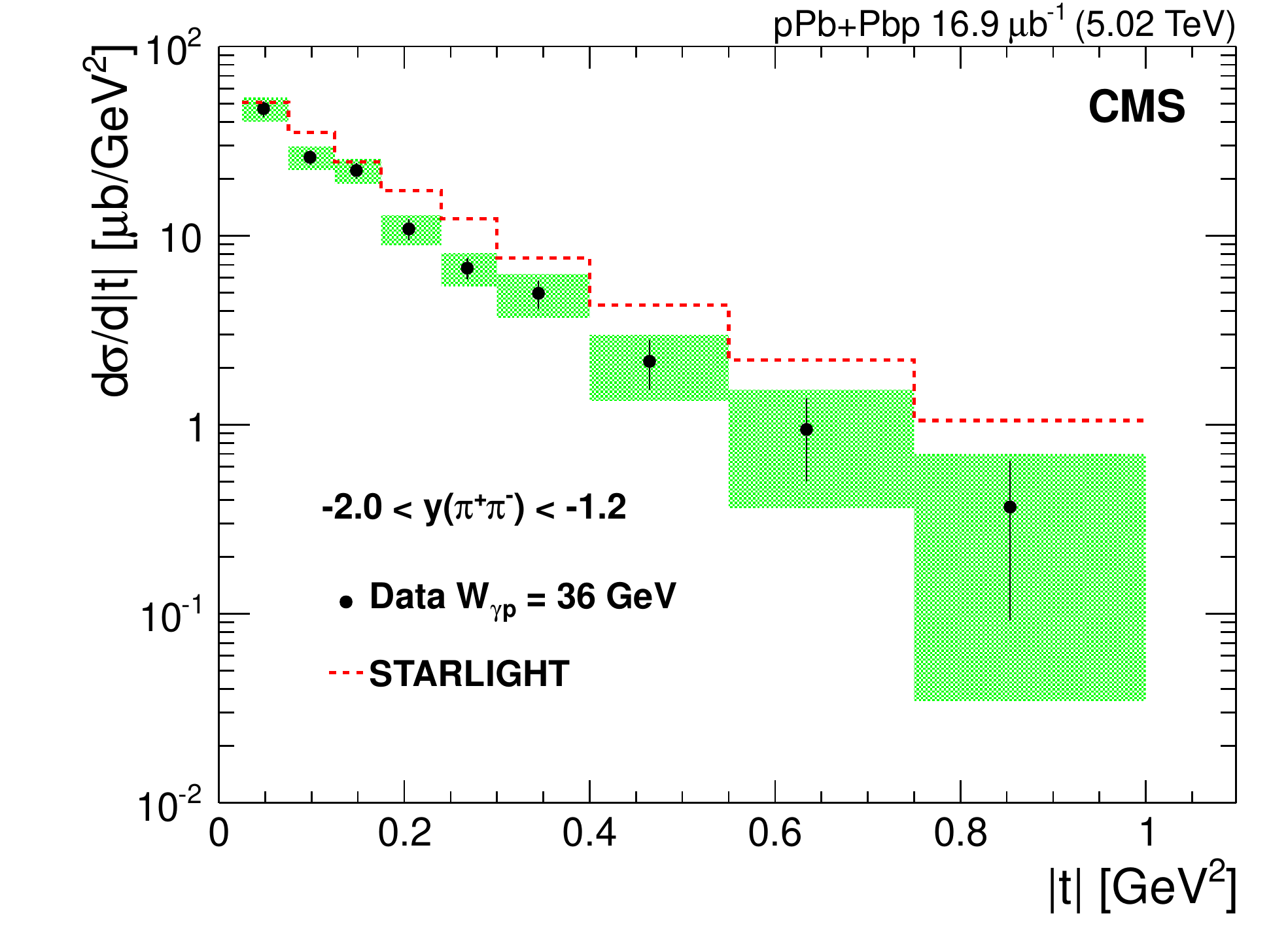}
\quad
\includegraphics[width=.45\textwidth]{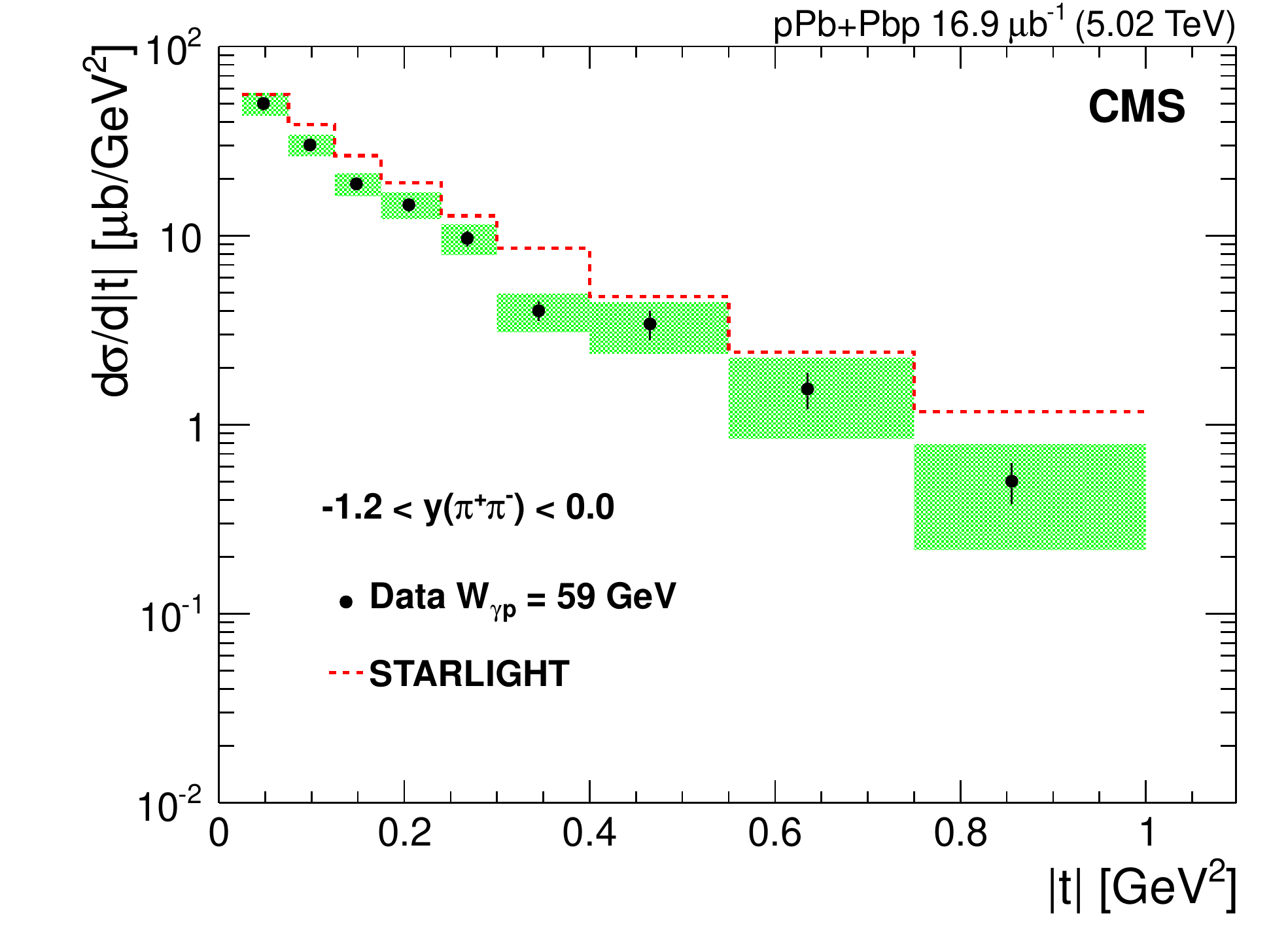}

\quad
\includegraphics[width=.45\textwidth]{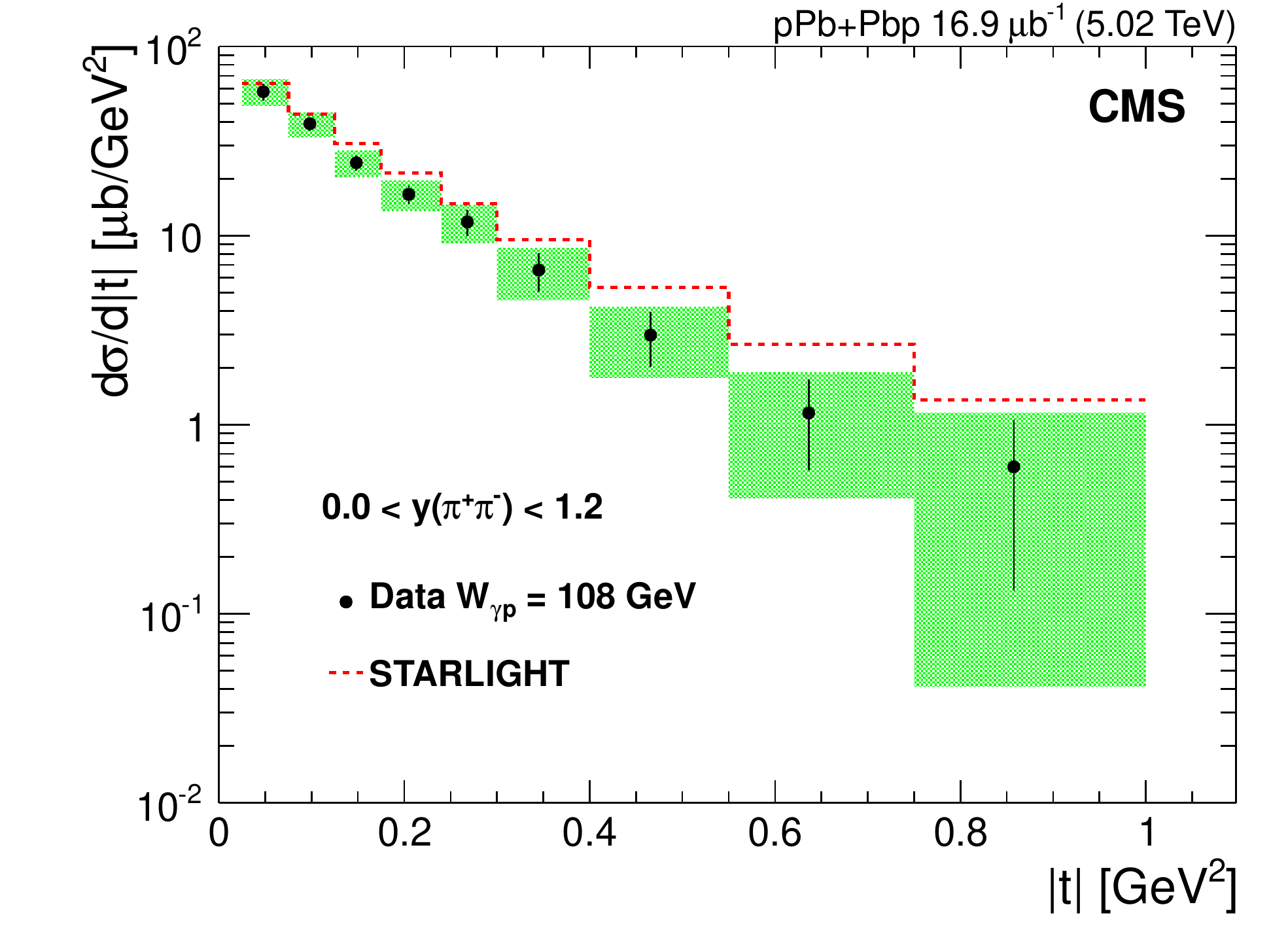}
\quad
\includegraphics[width=.45\textwidth]{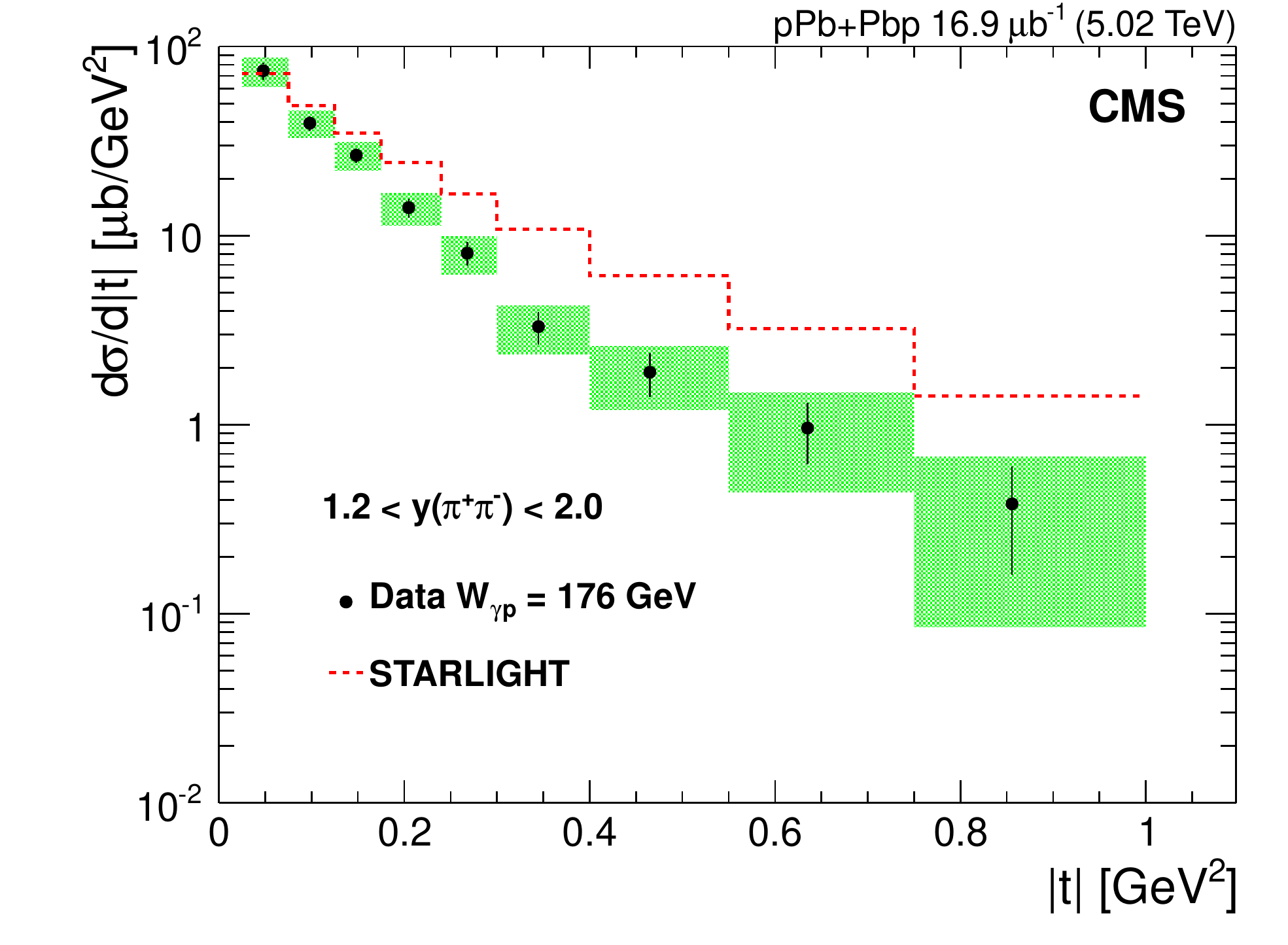}
 \caption[Differential cross section $\rd\sigma/\rd\abs{t}$ for exclusive $\Pgr^{0}$~photoproduction in four different rapidity
bins together with the\textsc{ starlight} prediction]{\label{figure:mass_res_SSL}
Differential cross section $\rd\sigma/\rd\abs{t}$ (full circles) in four different rapidity bins. The error bars show the statistical uncertainty,
whereas the shaded areas represent the statistical and systematic uncertainties added in quadrature. The dashed lines show the unweighted\textsc{ starlight} predictions.
}
\end{figure*}

The unfolded invariant mass distribution is studied in different $\abs{t}$ bins, and the extraction of the $\Pgr^{0}$~photoproduction cross section is performed in each bin. In order to compare with the HERA results, the \pt-related measurements are presented in terms of $\abs{t}$, which is
approximated as
$\abs{t} \approx (\pt^{\Pgpp\Pgpm})^{2}$. Figure~\ref{figure:mass_res_SSL} shows the differential cross sections as a function of $\abs{t}$, together with the unweighted\textsc{ starlight} prediction, whose slope parameter is independent of $W_{\Pgg\Pp}$. The\textsc{ starlight} prediction is systematically higher than the data in the high-$\abs{t}$ region.
This trend becomes more significant as $W_{\Pgg\Pp}$ increases.

\begin{figure*}[ht!]
\centering
\includegraphics[width=.8\textwidth]{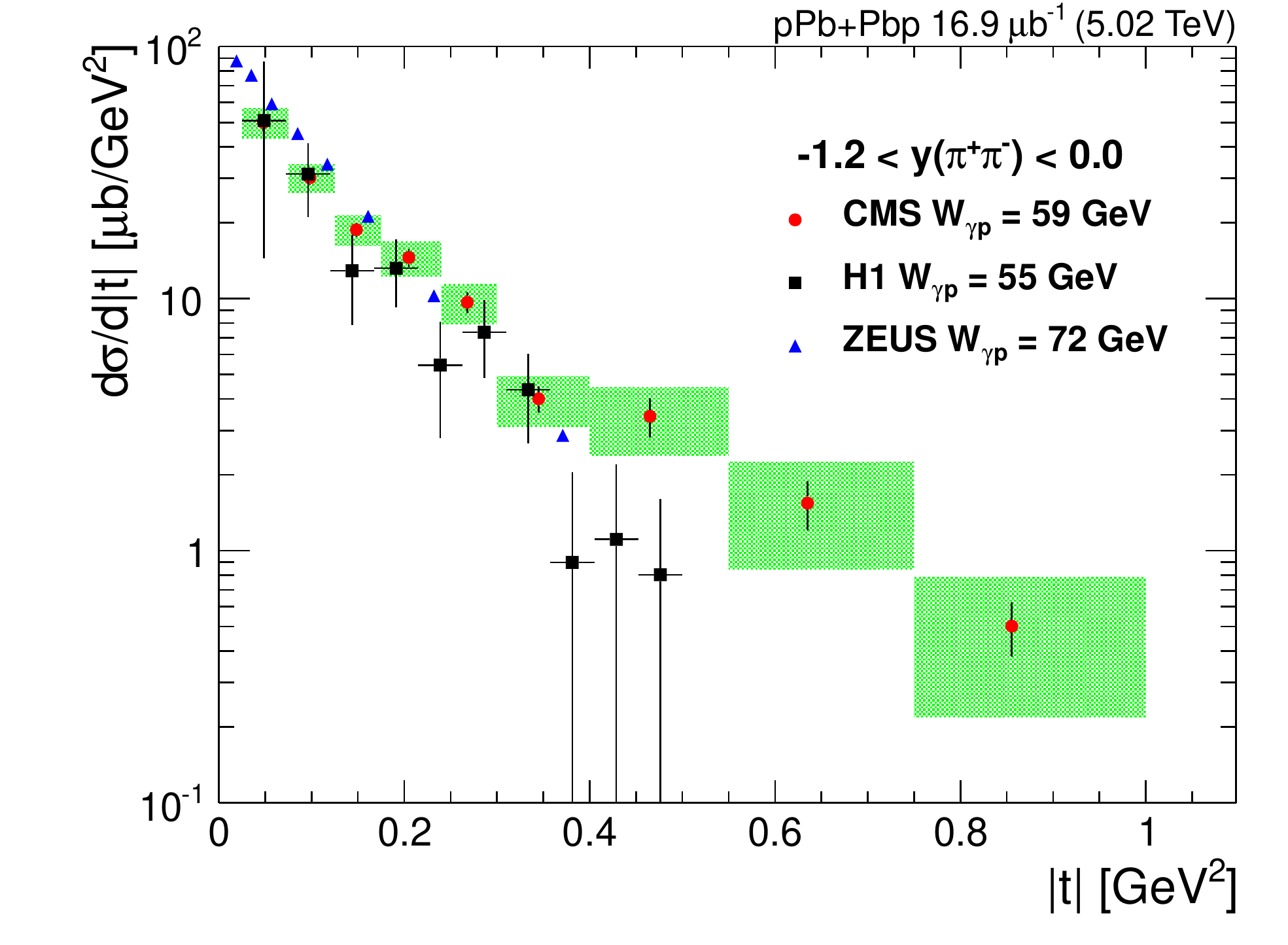}
   \caption[Differential cross section $\rd\sigma/\rd\abs{t}$ for exclusive $\Pgr^{0}$~photoproduction in the rapidity interval
$-1.2 < y(\Pgpp\Pgpm) < 0$ together with H1 and ZEUS data]{\label{figure:mass_S_HZ} Differential cross section
$\rd\sigma /\rd\abs{t}$ (full circles) for exclusive $\Pgr^{0}$~photoproduction in the rapidity interval $-1.2 < y_{\Pgpp\Pgpm} < 0$.
The square symbols indicate the H1 results, and the triangles the ZEUS results. The error bars show the statistical uncertainty,
while the shaded areas represent the statistical and systematic uncertainties added in quadrature.
For the H1 data~\cite{Aid:1996bs}, the error bars represent the statistical and systematic uncertainties added in quadrature, and for the ZEUS data~\cite{Breitweg:1997ed} the reported uncertainties are negligible.
}
\end{figure*}

Figure~\ref{figure:mass_S_HZ} shows the differential cross section $\rd\sigma/\rd\abs{t}$ in the rapidity interval $-1.2 < y(\Pgpp\Pgpm) < 0$ compared with the H1 and ZEUS results~\cite{Aid:1996bs,Breitweg:1997ed} in a similar $W_{\Pgg\Pp}$ range.
\begin{figure*}[ht!]
\centering
\includegraphics[width=.8\textwidth]{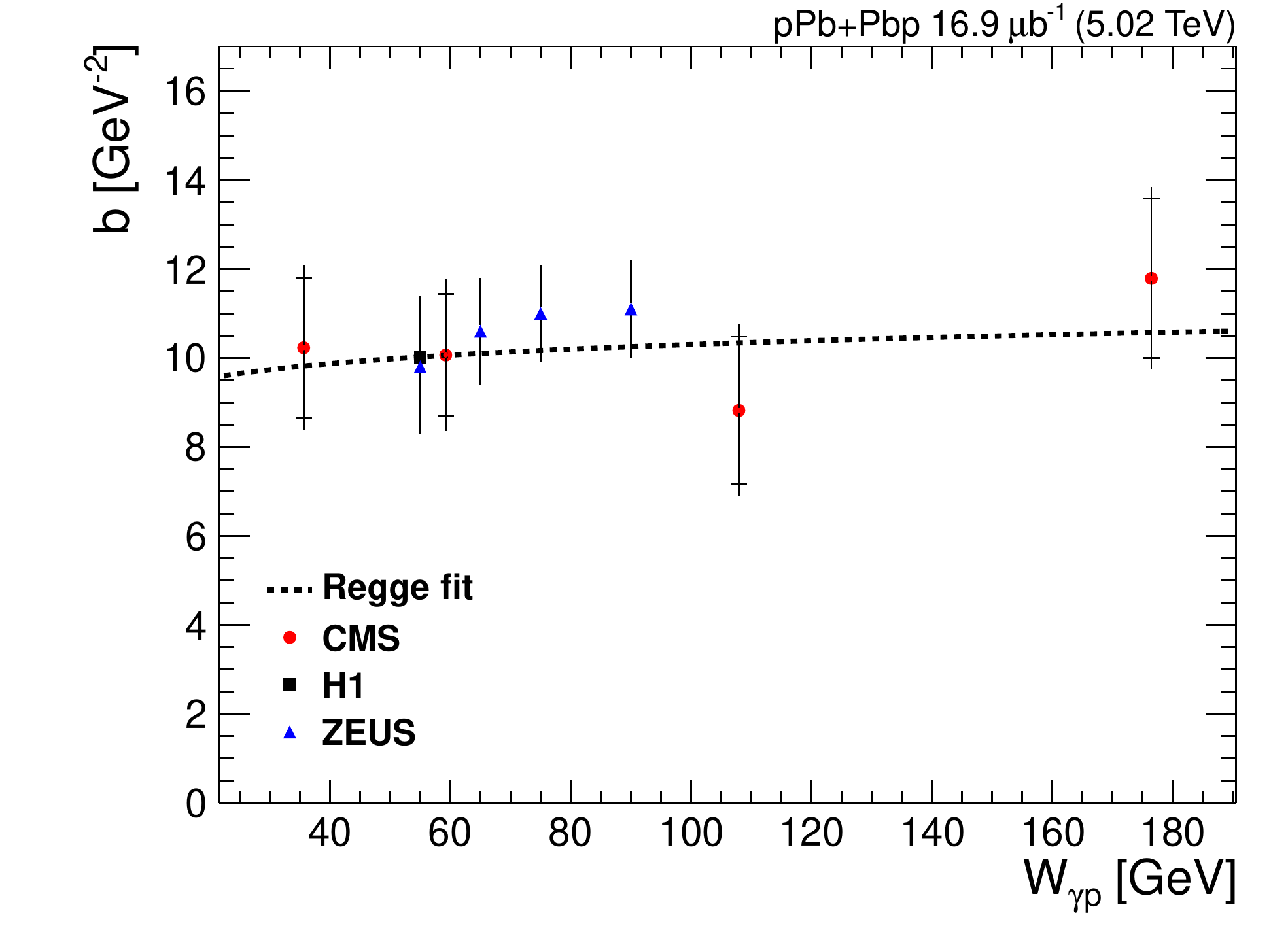}
 \caption[The slope parameter $b$ as a function of $W_{\Pgg\Pp}$]{\label{figure:b2_wgp}
The slope parameter $b$ extracted from the exponential fits of the differential cross sections $\rd\sigma/\rd\abs{t}$ shown as a function of $W_{\Pgg\Pp}$.
The inner error bars show the statistical uncertainty, while the outer error bars indicate the statistical and systematic uncertainties added in quadrature.
The dashed line shows the result of the Regge fit discussed in the text.
}
\end{figure*}

The differential cross section as a function of $\abs{t}$ is fitted with the form $A\re^{-bt+ct{^2}}$ in the region $0.025<\abs{t}<0.5$ $\GeVns^{2}$. For the integrated rapidity bin the fit gives $b=9.2\pm0.7$\stat$\GeVns^{-2}$ and $c=4.6\pm1.6$\stat$\GeVns^{-4}$. The resulting values of the slope $b$ are shown in Fig.~\ref{figure:b2_wgp} as a function of $W_{\Pgg\Pp}$, together with those measured by H1 and ZEUS~\cite{Aid:1996bs,Breitweg:1997ed}. The values of the parameter $c$ are found to be constant within the fit uncertainties. The Regge formula~\cite{Goulianos:1982vk} $b = b_{0} + 2\alpha'\ln(W_{\Pgg\Pp}/W_{0})^{2}$, which parametrizes the dependence of $b$ on the collision energy, is fitted to
the data using $W_{0} = 92.6$\GeV, the average centre-of-mass energy of the present data. The fit to the CMS data alone gives a pomeron slope of $\alpha' = 0.28 \pm 0.11\stat \pm 0.12$\syst$\GeVns^{-2}$, consistent with the ZEUS~\cite{Breitweg:1997ed} value and the Regge expectation of $0.25~\GeVns^{-2}$.

\begin{figure*}[ht!]
\includegraphics[width=.95\textwidth]{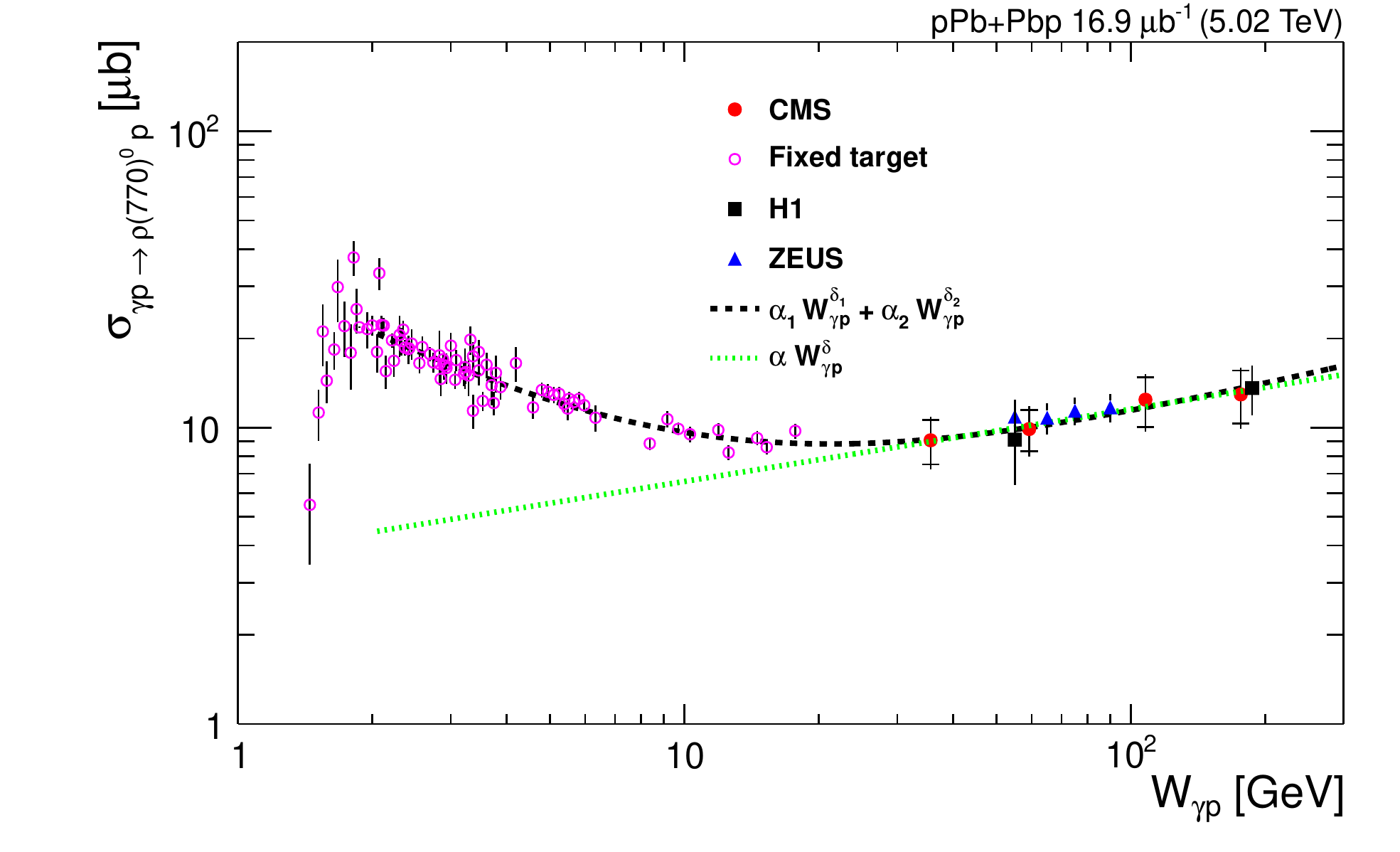}
\centering
 \caption[The total cross section $\sigma$ of $\Pgr^{0}$ photoproduction as a function of
$W_{\Pgg\Pp}$]{\label{figure:sigma_wgp1} Exclusive $\Pgr^{0}$~photoproduction cross section as a function of
$W_{\Pgg\Pp}$.
The inner bars show the statistical uncertainty,
while the outer bars represent the statistical and systematic uncertainties added in quadrature.
Fixed-target~\cite{fixed1,fixed2,fixed3,fixed4} and HERA~\cite{Aid:1996bs,Breitweg:1997ed} data are also shown.
The dashed lines indicate the results of the fits described in the text.
}
\end{figure*}

The resulting photon-proton cross section, obtained for $W_{\Pgg\Pp}$ between 29 and 213\GeV ($\langle W_{\Pgg\Pp} \rangle$ = 92.6\GeV) is extrapolated to the range $0<\abs{t}<0.5~\GeVns^{2}$ using the exponential fits just discussed and the\textsc{ starlight} predictions in order to allow direct comparison with previous experiments. The resulting value is $\sigma$ = 11.0 $\pm$ 1.4 \stat $\pm$ 1.0\syst $\mu $b. The photon-proton cross section values, $\sigma(\Pgg\Pp\to \Pgr^{0}\Pp)$, for all rapidity bins are presented in Table~\ref{tab:xs} and Fig.~\ref{figure:sigma_wgp1}. Figure~\ref{figure:sigma_wgp1} also shows a compilation of fixed-target~\cite{fixed1,fixed2,fixed3,fixed4} and HERA results~\cite{Aid:1996bs,Breitweg:1997ed}. The results of two fits are shown in Fig.~\ref{figure:sigma_wgp1}. The dashed line indicates the result of a fit to all the plotted data with the formula $\sigma = \alpha_{1} W^{\delta_{1}}_{\Pgg\Pp} + \alpha_{2} W^{\delta_{2}}_{\Pgg\Pp}$ (see e.g.~\cite{Newman:2013ada,Favart:2015umi}). The fit describes the data well and yields the values $\delta_{1}$ = $-$0.81 $\pm$ 0.04\stat $\pm$ 0.09\syst, $\delta_{2}$ = 0.36 $\pm$ 0.07\stat $\pm$ 0.05\syst. The CMS and HERA data are also fitted with the function $\sigma = \alpha W^{\delta}_{\Pgg\Pp}$ as shown in Fig.~\ref{figure:sigma_wgp1}. The fit yields $\delta$ = 0.24 $\pm$ 0.13\stat $\pm$ 0.04\syst.
Only statistical and uncorrelated systematic uncertainties are considered in these fits.

\section{Summary}
\label{sec:con}

The CMS Collaboration has made the first measurement of exclusive $\Pgr^{0}$~photoproduction off protons in ultraperipheral pPb collisions at $\sqrtsNN = 5.02\TeV$. The cross section for this process is measured in the photon-proton centre-of-mass energy interval
$29 < W_{\Pgg\Pp} <  213\GeV$. The results are consistent with those of the H1 and ZEUS Collaborations at
HERA, indicating that ion-proton collisions can be used in the same way as
electron-proton ones, with ions acting as a source of quasi-real photons.
 The combination of the present data and the earlier, lower energy results agrees with theory-inspired fits. The differential cross section $\rd\sigma/\rd\abs{t}$ for $\Pgr^{0}$ photoproduction is measured as a function of $W_{\Pgg\Pp}$. The\textsc{ starlight} prediction is systematically higher than the data in the high-$\abs{t}$ region.
This trend becomes more significant as $W_{\Pgg\Pp}$ increases.

\begin{acknowledgments}
We congratulate our colleagues in the CERN accelerator departments for the excellent performance of the LHC and thank the technical and administrative staffs at CERN and at other CMS institutes for their contributions to the success of the CMS effort. In addition, we gratefully acknowledge the computing centres and personnel of the Worldwide LHC Computing Grid for delivering so effectively the computing infrastructure essential to our analyses. Finally, we acknowledge the enduring support for the construction and operation of the LHC and the CMS detector provided by the following funding agencies: BMWFW and FWF (Austria); FNRS and FWO (Belgium); CNPq, CAPES, FAPERJ, and FAPESP (Brazil); MES (Bulgaria); CERN; CAS, MoST, and NSFC (China); COLCIENCIAS (Colombia); MSES and CSF (Croatia); RPF (Cyprus); SENESCYT (Ecuador); MoER, ERC IUT and ERDF (Estonia); Academy of Finland, MEC, and HIP (Finland); CEA and CNRS/IN2P3 (France); BMBF, DFG, and HGF (Germany); GSRT (Greece); OTKA and NIH (Hungary); DAE and DST (India); IPM (Iran); SFI (Ireland); INFN (Italy); MSIP and NRF (Republic of Korea); LAS (Lithuania); MOE and UM (Malaysia); BUAP, CINVESTAV, CONACYT, LNS, SEP, and UASLP-FAI (Mexico); MBIE (New Zealand); PAEC (Pakistan); MSHE and NSC (Poland); FCT (Portugal); JINR (Dubna); MON, RosAtom, RAS and RFBR (Russia); MESTD (Serbia); SEIDI and CPAN (Spain); Swiss Funding Agencies (Switzerland); MST (Taipei); ThEPCenter, IPST, STAR and NSTDA (Thailand); TUBITAK and TAEK (Turkey); NASU and SFFR (Ukraine); STFC (United Kingdom); DOE and NSF (USA).

\hyphenation{Rachada-pisek} Individuals have received support from the Marie-Curie programme and the European Research Council and Horizon 2020 Grant, contract No. 675440 (European Union); the Leventis Foundation; the A.P.\ Sloan Foundation; the Alexander von Humboldt Foundation; the Belgian Federal Science Policy Office; the Fonds pour la Formation \`a la Recherche dans l'Industrie et dans l'Agriculture (FRIA-Belgium); the Agentschap voor Innovatie door Wetenschap en Technologie (IWT-Belgium); the F.R.S.-FNRS and FWO (Belgium) under the ``Excellence of Science -- EOS" -- be.h project n.\ 30820817; the Ministry of Education, Youth and Sports (MEYS) of the Czech Republic; the Lend\"ulet (``Momentum") Programme and the J\'anos Bolyai Research Scholarship of the Hungarian Academy of Sciences, the New National Excellence Program \'UNKP, the NKFIA research grants 123842, 123959, 124845, 124850, and 125105 (Hungary); the Council of Science and Industrial Research, India; the HOMING PLUS programme of the Foundation for Polish Science, cofinanced from European Union, Regional Development Fund, the Mobility Plus programme of the Ministry of Science and Higher Education, the National Science Center (Poland), contracts Harmonia 2014/14/M/ST2/00428, Opus 2014/13/B/ST2/02543, 2014/15/B/ST2/03998, and 2015/19/B/ST2/02861, Sonata-bis 2012/07/E/ST2/01406; the National Priorities Research Program by Qatar National Research Fund; the Programa Estatal de Fomento de la Investigaci{\'o}n Cient{\'i}fica y T{\'e}cnica de Excelencia Mar\'{\i}a de Maeztu, grant MDM-2015-0509 and the Programa Severo Ochoa del Principado de Asturias; the Thalis and Aristeia programmes cofinanced by EU-ESF and the Greek NSRF; the Rachadapisek Sompot Fund for Postdoctoral Fellowship, Chulalongkorn University and the Chulalongkorn Academic into Its 2nd Century Project Advancement Project (Thailand); the Welch Foundation, contract C-1845; and the Weston Havens Foundation (USA).
\end{acknowledgments}
\clearpage
\bibliography{auto_generated}

\cleardoublepage \appendix\section{The CMS Collaboration \label{app:collab}}\begin{sloppypar}\hyphenpenalty=5000\widowpenalty=500\clubpenalty=5000\input{FSQ-16-007-authorlist.tex}\end{sloppypar}
\end{document}

%% file: FSQ-16-007-authorlist.tex
\vskip\cmsinstskip
\textbf{Yerevan Physics Institute, Yerevan, Armenia}\\*[0pt]
A.M.~Sirunyan, A.~Tumasyan
\vskip\cmsinstskip
\textbf{Institut f\"{u}r Hochenergiephysik, Wien, Austria}\\*[0pt]
W.~Adam, F.~Ambrogi, E.~Asilar, T.~Bergauer, J.~Brandstetter, M.~Dragicevic, J.~Er\"{o}, A.~Escalante~Del~Valle, M.~Flechl, R.~Fr\"{u}hwirth\cmsAuthorMark{1}, V.M.~Ghete, J.~Hrubec, M.~Jeitler\cmsAuthorMark{1}, N.~Krammer, I.~Kr\"{a}tschmer, D.~Liko, T.~Madlener, I.~Mikulec, N.~Rad, H.~Rohringer, J.~Schieck\cmsAuthorMark{1}, R.~Sch\"{o}fbeck, M.~Spanring, D.~Spitzbart, A.~Taurok, W.~Waltenberger, J.~Wittmann, C.-E.~Wulz\cmsAuthorMark{1}, M.~Zarucki
% ...
\vskip\cmsinstskip
% ...
1:  Also at Vienna University of Technology, Vienna, Austria\\
% ...


%% file: FSQ-16-007_temp.bbl
\providecommand{\href}[2]{#2}\begingroup\raggedright\begin{thebibliography}{10}%
\makeatletter
\providecommand{\hrefCMSnoop }[0]{\@secondoftwo}%
\makeatother
\providecommand{\doi}{\texttt{doi:}\begingroup \urlstyle{tt}\Url}

\bibitem{Baltz:2007kq}
\hrefCMSnoop {}{A.~J. Baltz, ``The physics of ultraperipheral collisions at the
  {LHC}'',} \textit{ Phys. Rept.} \textbf{ 458} (2008) 1,
  \href{http://dx.doi.org/10.1016/j.physrep.2007.12.001}{\doi{10.1016/j.physrep.2007.12.001}},
\href{http://www.arXiv.org/abs/0706.3356}{\texttt{arXiv:0706.3356}}.

\bibitem{Contreras:2015dqa}
\hrefCMSnoop {}{J.~G. Contreras and J.~D. Tapia~Takaki, ``Ultra-peripheral
  heavy-ion collisions at the {LHC}'',} \textit{ Int. J. Mod. Phys. A} \textbf{
  30} (2015) 1542012,
\href{http://dx.doi.org/10.1142/S0217751X15420129}{\doi{10.1142/S0217751X15420129}}.

\bibitem{Frankfurt:2006tp}
\hrefCMSnoop {}{L.~Frankfurt, M.~Strikman, and M.~Zhalov, ``Elastic and large t
  rapidity gap vector meson production in ultraperipheral proton-ion
  collisions'',} \textit{ Phys. Lett. B} \textbf{ 640} (2006) 162,
  \href{http://dx.doi.org/10.1016/j.physletb.2006.07.059}{\doi{10.1016/j.physletb.2006.07.059}},
\href{http://www.arXiv.org/abs/hep-ph/0605160}{\texttt{arXiv:hep-ph/0605160}}.

\bibitem{Guzey:2013taa}
\hrefCMSnoop {}{V.~Guzey and M.~Zhalov, ``Rapidity and momentum transfer
  distributions of coherent $\mathrm{J/}\psi$ photoproduction in
  ultraperipheral {pPb} collisions at the {LHC}'',} \textit{ JHEP} \textbf{ 02}
  (2014) 046,
  \href{http://dx.doi.org/10.1007/JHEP02(2014)046}{\doi{10.1007/JHEP02(2014)046}},
\href{http://www.arXiv.org/abs/1307.6689}{\texttt{arXiv:1307.6689}}.

\bibitem{Armesto:2014sma}
\hrefCMSnoop {}{N.~Armesto and A.~H. Rezaeian, ``Exclusive vector meson
  production at high energies and gluon saturation'',} \textit{ Phys. Rev. D}
  \textbf{ 90} (2014) 054003,
  \href{http://dx.doi.org/10.1103/PhysRevD.90.054003}{\doi{10.1103/PhysRevD.90.054003}},
\href{http://www.arXiv.org/abs/1402.4831}{\texttt{arXiv:1402.4831}}.

\bibitem{Toll:2012mb}
\hrefCMSnoop {}{T.~Toll and T.~Ullrich, ``Exclusive diffractive processes in
  electron-ion collisions'',} \textit{ Phys. Rev. C} \textbf{ 87} (2013)
  024913,
  \href{http://dx.doi.org/10.1103/PhysRevC.87.024913}{\doi{10.1103/PhysRevC.87.024913}},
\href{http://www.arXiv.org/abs/1211.3048}{\texttt{arXiv:1211.3048}}.

\bibitem{Jones:2013pga}
\hrefCMSnoop {}{S.~P. Jones, A.~D. Martin, M.~G. Ryskin, and T.~Teubner,
  ``{Probes of the small $x$ gluon via exclusive $\mathrm{J/}\psi$ and
  $\Upsilon$ production at {HERA} and the {LHC}}'',} \textit{ JHEP} \textbf{
  11} (2013) 085,
  \href{http://dx.doi.org/10.1007/JHEP11(2013)085}{\doi{10.1007/JHEP11(2013)085}},
\href{http://www.arXiv.org/abs/1307.7099}{\texttt{arXiv:1307.7099}}.

\bibitem{Goncalves:2018blz}
\hrefCMSnoop {}{V.~P. Goncalves, F.~S. Navarra, and D.~Spiering, ``{Exclusive
  \Pgr and $J/\Psi$ photoproduction in ultraperipheral $pA$ collisions:
  Predictions of the gluon saturation models for the momentum transfer
  distributions}'',} \textit{ Phys. Lett. B} \textbf{ 791} (2019) 299,
  \href{http://dx.doi.org/10.1016/j.physletb.2019.03.007}{\doi{10.1016/j.physletb.2019.03.007}},
\href{http://www.arXiv.org/abs/1811.09124}{\texttt{arXiv:1811.09124}}.

\bibitem{Cepila:2018zky}
\hrefCMSnoop {}{J.~Cepila, J.~G. Contreras, M.~Krelina, and J.~D. Tapia~Takaki,
  ``{Mass dependence of vector meson photoproduction off protons and nuclei
  within the energy-dependent hot-spot model}'',} \textit{ Nucl. Phys. B}
  \textbf{ 934} (2018) 330,
  \href{http://dx.doi.org/10.1016/j.nuclphysb.2018.07.010}{\doi{10.1016/j.nuclphysb.2018.07.010}},
\href{http://www.arXiv.org/abs/1804.05508}{\texttt{arXiv:1804.05508}}.

\bibitem{Cepila:2016uku}
\hrefCMSnoop {}{J.~Cepila, J.~G. Contreras, and J.~D. Tapia~Takaki, ``{Energy
  dependence of dissociative $\mathrm{J/}\psi$ photoproduction as a signature
  of gluon saturation at the LHC}'',} \textit{ Phys. Lett. B} \textbf{ 766}
  (2017) 186,
  \href{http://dx.doi.org/10.1016/j.physletb.2016.12.063}{\doi{10.1016/j.physletb.2016.12.063}},
\href{http://www.arXiv.org/abs/1608.07559}{\texttt{arXiv:1608.07559}}.

\bibitem{TheALICE:2014dwa}
\hrefCMSnoop {}{{{ALICE}} Collaboration, ``Exclusive $\mathrm{J/}\psi$
  photoproduction off protons in ultraperipheral {p-Pb} collisions at
  $\sqrt{s_{\rm NN}}=5.02~\mathrm{TeV}$'',} \textit{ Phys. Rev. Lett.} \textbf{
  113} (2014) 232504,
  \href{http://dx.doi.org/10.1103/PhysRevLett.113.232504}{\doi{10.1103/PhysRevLett.113.232504}},
\href{http://www.arXiv.org/abs/1406.7819}{\texttt{arXiv:1406.7819}}.

\bibitem{Abelev:2012ba}
\hrefCMSnoop {}{{ALICE Collaboration}, ``Coherent $\mathrm{J/}\psi$
  photoproduction in ultra-peripheral {Pb-Pb} collisions at $\sqrt{s_{NN}} =
  2.76~\mathrm{TeV}$'',} \textit{ Phys. Lett. B} \textbf{ 718} (2013) 1273,
  \href{http://dx.doi.org/10.1016/j.physletb.2012.11.059}{\doi{10.1016/j.physletb.2012.11.059}},
\href{http://www.arXiv.org/abs/1209.3715}{\texttt{arXiv:1209.3715}}.

\bibitem{Aaij:2014iea}
\hrefCMSnoop {}{{{LHCb}} Collaboration, ``Updated measurements of exclusive
  $\mathrm{J/}\psi$ and $\psi$(2s) production cross-sections in pp collisions
  at $\sqrt{s}=7~\mathrm{TeV}$'',} \textit{ J. Phys. G} \textbf{ 41} (2014)
  055002,
  \href{http://dx.doi.org/10.1088/0954-3899/41/5/055002}{\doi{10.1088/0954-3899/41/5/055002}},
\href{http://www.arXiv.org/abs/1401.3288}{\texttt{arXiv:1401.3288}}.

\bibitem{Aaij:2015kea}
\hrefCMSnoop {}{{{LHCb}} Collaboration, ``Measurement of the exclusive {Y}
  production cross-section in pp collisions at
  $\sqrt{s}=7~\mathrm{TeV}~\mathrm{and}~8~\mathrm{TeV}$'',} \textit{ JHEP}
  \textbf{ 09} (2015) 084,
  \href{http://dx.doi.org/10.1007/JHEP09(2015)084}{\doi{10.1007/JHEP09(2015)084}},
\href{http://www.arXiv.org/abs/1505.08139}{\texttt{arXiv:1505.08139}}.

\bibitem{Bauer:1977iq}
\hrefCMSnoop {}{T.~H. Bauer, R.~D. Spital, D.~R. Yennie, and F.~M. Pipkin,
  ``The hadronic properties of the photon in high-energy interactions'',}
  \textit{ Rev. Mod. Phys.} \textbf{ 50} (1978) 261,
  \href{http://dx.doi.org/10.1103/RevModPhys.50.261}{\doi{10.1103/RevModPhys.50.261}}.
[Erratum: \DOI{10.1103/RevModPhys.51.407}].

\bibitem{Crittenden:1997yz}
\hrefCMSnoop {}{J.~A. Crittenden, ``Exclusive production of neutral vector
  mesons at the electron - proton collider {HERA}''}.
\newblock PhD thesis, {Physikalisches Institut der Universitaet Bonn}, 1997.
\newblock
  \href{http://www.arXiv.org/abs/hep-ex/9704009}{\texttt{arXiv:hep-ex/9704009}}.
\newblock
DESY-97-068, BONN-IR-97-01.

\bibitem{Breitweg:1997ed}
\hrefCMSnoop {}{{{ZEUS}} Collaboration, ``Elastic and proton dissociative
  $\rho^0$ photoproduction at {HERA}'',} \textit{ Eur. Phys. J. C} \textbf{ 2}
  (1998) 247,
  \href{http://dx.doi.org/10.1007/s100529800834}{\doi{10.1007/s100529800834}},
\href{http://www.arXiv.org/abs/hep-ex/9712020}{\texttt{arXiv:hep-ex/9712020}}.

\bibitem{Aid:1996bs}
\hrefCMSnoop {}{{{H1}} Collaboration, ``Elastic photoproduction of $\rho^0$
  mesons at {HERA}'',} \textit{ Nucl. Phys. B} \textbf{ 463} (1996) 3,
  \href{http://dx.doi.org/10.1016/0550-3213(96)00045-4}{\doi{10.1016/0550-3213(96)00045-4}},
\href{http://www.arXiv.org/abs/hep-ex/9601004}{\texttt{arXiv:hep-ex/9601004}}.

\bibitem{Newman:2013ada}
\hrefCMSnoop {}{P.~Newman and M.~Wing, ``The hadronic final state at {HERA}'',}
  \textit{ Rev. Mod. Phys.} \textbf{ 86} (2014) 1037,
  \href{http://dx.doi.org/10.1103/RevModPhys.86.1037}{\doi{10.1103/RevModPhys.86.1037}},
\href{http://www.arXiv.org/abs/1308.3368}{\texttt{arXiv:1308.3368}}.

\bibitem{Favart:2015umi}
\hrefCMSnoop {}{L.~Favart, M.~Guidal, T.~Horn, and P.~Kroll, ``Deeply virtual
  meson production on the nucleon'',} \textit{ Eur. Phys. J. A} \textbf{ 52}
  (2016) 158,
  \href{http://dx.doi.org/10.1140/epja/i2016-16158-2}{\doi{10.1140/epja/i2016-16158-2}},
\href{http://www.arXiv.org/abs/1511.04535}{\texttt{arXiv:1511.04535}}.

\bibitem{Adler:2002sc}
\hrefCMSnoop {}{{{STAR}} Collaboration, ``Coherent $\rho^0$ production in
  ultraperipheral heavy ion collisions'',} \textit{ Phys. Rev. Lett.} \textbf{
  89} (2002) 272302,
  \href{http://dx.doi.org/10.1103/PhysRevLett.89.272302}{\doi{10.1103/PhysRevLett.89.272302}},
\href{http://www.arXiv.org/abs/nucl-ex/0206004}{\texttt{arXiv:nucl-ex/0206004}}.

\bibitem{Abelev:2007nb}
\hrefCMSnoop {}{{{STAR}} Collaboration, ``$\rho^0$ photoproduction in
  ultraperipheral relativistic heavy ion collisions at $\sqrt{s_{\mathrm{NN}}}
  = 200~\mathrm{GeV}$'',} \textit{ Phys. Rev. C} \textbf{ 77} (2008) 034910,
  \href{http://dx.doi.org/10.1103/PhysRevC.77.034910}{\doi{10.1103/PhysRevC.77.034910}},
\href{http://www.arXiv.org/abs/0712.3320}{\texttt{arXiv:0712.3320}}.

\bibitem{Adamczyk:2017vfu}
\hrefCMSnoop {}{{{STAR}} Collaboration, ``Coherent diffractive photoproduction
  of $\rho^0$ mesons on gold nuclei at $200~\mathrm{GeV}$/nucleon-pair at the
  relativistic heavy ion collider'',} \textit{ Phys. Rev. C} \textbf{ 96}
  (2017) 054904,
  \href{http://dx.doi.org/10.1103/PhysRevC.96.054904}{\doi{10.1103/PhysRevC.96.054904}},
\href{http://www.arXiv.org/abs/1702.07705}{\texttt{arXiv:1702.07705}}.

\bibitem{Adam:2015gsa}
\hrefCMSnoop {}{{{ALICE}} Collaboration, ``Coherent $\rho^{0}$ photoproduction
  in ultra-peripheral {Pb-Pb} collisions at $
  \sqrt{s_{\mathrm{NN}}}=2.76~\mathrm{TeV}$'',} \textit{ JHEP} \textbf{ 09}
  (2015) 095,
  \href{http://dx.doi.org/10.1007/JHEP09(2015)095}{\doi{10.1007/JHEP09(2015)095}},
\href{http://www.arXiv.org/abs/1503.09177}{\texttt{arXiv:1503.09177}}.

\bibitem{Frankfurt:2015cwa}
\hrefCMSnoop {}{L.~Frankfurt, V.~Guzey, M.~Strikman, and M.~Zhalov, ``{Nuclear
  shadowing in photoproduction of $\rho(770)^{0}$ mesons in ultraperipheral
  nucleus collisions at RHIC and the LHC}'',} \textit{ Phys. Lett. B} \textbf{
  752} (2016) 51,
  \href{http://dx.doi.org/10.1016/j.physletb.2015.11.012}{\doi{10.1016/j.physletb.2015.11.012}},
\href{http://www.arXiv.org/abs/1506.07150}{\texttt{arXiv:1506.07150}}.

\bibitem{Chatrchyan:2014fea}
\hrefCMSnoop {}{{CMS Collaboration}, ``Description and performance of track and
  primary-vertex reconstruction with the {CMS} tracker'',} \textit{ JINST}
  \textbf{ 9} (2014) P10009,
  \href{http://dx.doi.org/10.1088/1748-0221/9/10/P10009}{\doi{10.1088/1748-0221/9/10/P10009}},
\href{http://www.arXiv.org/abs/1405.6569}{\texttt{arXiv:1405.6569}}.

\bibitem{Chatrchyan:2008zzk}
\hrefCMSnoop {}{{CMS Collaboration}, ``The {CMS} experiment at the {CERN}
  {LHC}'',} \textit{ JINST} \textbf{ 3} (2008) S08004,
\href{http://dx.doi.org/10.1088/1748-0221/3/08/S08004}{\doi{10.1088/1748-0221/3/08/S08004}}.

\bibitem{Klein:2016yzr}
S.~R. Klein\hrefCMSnoop {}{ {et~al.}, ``{STARLIGHT}: A {Monte Carlo} simulation
  program for ultra-peripheral collisions of relativistic ions'',} \textit{
  Comput. Phys. Commun.} \textbf{ 212} (2017) 258,
  \href{http://dx.doi.org/10.1016/j.cpc.2016.10.016}{\doi{10.1016/j.cpc.2016.10.016}},
\href{http://www.arXiv.org/abs/1607.03838}{\texttt{arXiv:1607.03838}}.

\bibitem{Agostinelli:2002hh}
\hrefCMSnoop {}{{{GEANT4}} Collaboration, ``{\GEANTfour}---a simulation
  toolkit'',} \textit{ Nucl. Instrum. Meth. A} \textbf{ 506} (2003) 250,
\href{http://dx.doi.org/10.1016/S0168-9002(03)01368-8}{\doi{10.1016/S0168-9002(03)01368-8}}.

\bibitem{Khachatryan:2016bia}
\hrefCMSnoop {}{{CMS Collaboration}, ``{The CMS trigger system}'',} \textit{
  JINST} \textbf{ 12} (2017) P01020,
  \href{http://dx.doi.org/10.1088/1748-0221/12/01/P01020}{\doi{10.1088/1748-0221/12/01/P01020}},
\href{http://www.arXiv.org/abs/1609.02366}{\texttt{arXiv:1609.02366}}.

\bibitem{Khachatryan:2010pw}
\hrefCMSnoop {}{{{CMS}} Collaboration, ``{CMS} tracking performance results
  from early {LHC} operation'',} \textit{ Eur. Phys. J. C} \textbf{ 70} (2010)
  1165,
  \href{http://dx.doi.org/10.1140/epjc/s10052-010-1491-3}{\doi{10.1140/epjc/s10052-010-1491-3}},
\href{http://www.arXiv.org/abs/1007.1988}{\texttt{arXiv:1007.1988}}.

\bibitem{Aaron:2009xp}
\hrefCMSnoop {}{{{H1}} Collaboration, ``Diffractive electroproduction of $\rho$
  and $\phi$ mesons at {HERA}'',} \textit{ JHEP} \textbf{ 05} (2010) 032,
  \href{http://dx.doi.org/10.1007/JHEP05(2010)032}{\doi{10.1007/JHEP05(2010)032}},
\href{http://www.arXiv.org/abs/0910.5831}{\texttt{arXiv:0910.5831}}.

\bibitem{Tanabashi:2018oca}
\hrefCMSnoop {}{{Particle Data Group}, M.~Tanabashi {et~al.}, ``Review of
  particle physics'',} \textit{ Phys. Rev. D} \textbf{ 98} (2018) 030001,
  \href{http://dx.doi.org/10.1103/PhysRevD.98.030001}{\doi{10.1103/PhysRevD.98.030001}}.

\bibitem{Abelev:2009aa}
\hrefCMSnoop {}{{STAR} Collaboration, ``{Observation of $\pi^{+} \pi^{-}
  \pi^{+} \pi^{-}$ photoproduction in ultraperipheral heavy-ion Collisions at
  STAR}'',} \textit{ Phys. Rev. C} \textbf{ 81} (2010) 044901,
  \href{http://dx.doi.org/10.1103/PhysRevC.81.044901}{\doi{10.1103/PhysRevC.81.044901}},
\href{http://www.arXiv.org/abs/0912.0604}{\texttt{arXiv:0912.0604}}.

\bibitem{Schramm:1996aa}
\hrefCMSnoop {}{A.~J. Schramm and D.~H. Reeves, ``Production of $\eta$ mesons
  in double {Pomeron} exchange'',} \textit{ Phys. Rev. D} \textbf{ 55} (1997)
  7312,
  \href{http://dx.doi.org/10.1103/PhysRevD.55.7312}{\doi{10.1103/PhysRevD.55.7312}},
\href{http://www.arXiv.org/abs/hep-ph/9611330}{\texttt{arXiv:hep-ph/9611330}}.

\bibitem{DAGOSTINI1995487}
\hrefCMSnoop {}{G.~D'Agostini, ``A multidimensional unfolding method based on
  {Bayes} theorem'',} \textit{ Nucl. Instrum. Meth. A} \textbf{ 362} (1995)
  487,
  \href{http://dx.doi.org/10.1016/0168-9002(95)00274-X}{\doi{10.1016/0168-9002(95)00274-X}}.

\bibitem{Adye:2011gm}
\hrefCMSnoop {}{T.~Adye, ``{Unfolding algorithms and tests using
  {R}oo{U}nfold}'',} in \textit{ {Proceedings, PHYSTAT 2011 Workshop on
  Statistical Issues Related to Discovery Claims in Search Experiments and
  Unfolding, CERN, Geneva, Switzerland 17-20 January 2011}}, p.~313, CERN.
\newblock CERN, Geneva, 2011.
\newblock \href{http://www.arXiv.org/abs/1105.1160}{\texttt{arXiv:1105.1160}}.
\newblock
\href{http://dx.doi.org/10.5170/CERN-2011-006.313}{\doi{10.5170/CERN-2011-006.313}}.

\bibitem{Mcclellan:1972cz}
G.~McClellan\hrefCMSnoop {}{ {et~al.}, ``Photoproduction of neutral $\rho^{0}$
  mesons'',} \textit{ Phys. Rev. D} \textbf{ 4} (1971) 2683,
\href{http://dx.doi.org/10.1103/PhysRevD.4.2683}{\doi{10.1103/PhysRevD.4.2683}}.

\bibitem{Soding:1965nh}
\hrefCMSnoop {}{P.~S{\"o}ding, ``On the apparent shift of the $\rho$ meson mass
  in photoproduction'',} \textit{ Phys. Lett.} \textbf{ 19} (1966) 702,
\href{http://dx.doi.org/10.1016/0031-9163(66)90451-3}{\doi{10.1016/0031-9163(66)90451-3}}.

\bibitem{Alvensleben:1971hz}
\hrefCMSnoop {}{H.~Alvensleben {et~al.}, ``{Precise determination of rho-omega
  interference parameters from photoproduction of vector mesons off nucleon and
  nuclei}'',} \textit{ Phys. Rev. Lett.} \textbf{ 27} (1971) 888,
\href{http://dx.doi.org/10.1103/PhysRevLett.27.888}{\doi{10.1103/PhysRevLett.27.888}}.

\bibitem{CMS:2014uoa}
\href {https://cds.cern.ch/record/1643269}{{CMS Collaboration}, ``Luminosity
  calibration for the 2013 proton-lead and proton-proton data taking'',} CMS
  Physics Analysis Summary CMS-PAS-LUM-13-002, 2013.

\bibitem{Sirunyan:2017nsj}
\hrefCMSnoop {}{{CMS Collaboration}, ``Measurement of the inclusive energy
  spectrum in the very forward direction in proton-proton collisions at
  $\sqrt{s}=13~\mathrm{TeV}$'',} \textit{ JHEP} \textbf{ 08} (2017) 046,
  \href{http://dx.doi.org/10.1007/JHEP08(2017)046}{\doi{10.1007/JHEP08(2017)046}},
\href{http://www.arXiv.org/abs/1701.08695}{\texttt{arXiv:1701.08695}}.

\bibitem{Ross:1965qa}
\hrefCMSnoop {}{M.~H. Ross and L.~Stodolsky, ``Photon dissociation model for
  vector meson photoproduction'',} \textit{ Phys. Rev.} \textbf{ 149} (1966)
  1172,
\href{http://dx.doi.org/10.1103/PhysRev.149.1172}{\doi{10.1103/PhysRev.149.1172}}.

\bibitem{Goulianos:1982vk}
\hrefCMSnoop {}{K.~A. Goulianos, ``Diffractive interactions of hadrons at high
  energies'',} \textit{ Phys. Rept.} \textbf{ 101} (1983) 169,
\href{http://dx.doi.org/10.1016/0370-1573(83)90010-8}{\doi{10.1016/0370-1573(83)90010-8}}.

\bibitem{fixed1}
\hrefCMSnoop {}{{E665} Collaboration, ``{Diffractive production of
  $\rho^0(770)$ mesons in muon-proton interactions at 470-GeV}'',} \textit{ Z.
  Phys. C} \textbf{ 74} (1997) 237,
\href{http://dx.doi.org/10.1007/s002880050386}{\doi{10.1007/s002880050386}}.

\bibitem{fixed2}
\hrefCMSnoop {}{D.~G. Cassel {et~al.}, ``{Exclusive $\rho^0$, $\omega$ and
  $\phi$ electroproduction}'',} \textit{ Phys. Rev. D} \textbf{ 24} (1981)
  2787,
\href{http://dx.doi.org/10.1103/PhysRevD.24.2787}{\doi{10.1103/PhysRevD.24.2787}}.

\bibitem{fixed3}
\hrefCMSnoop {}{{CLAS} Collaboration, ``{Exclusive $\rho^{0}$ meson
  electroproduction from hydrogen at {CLAS}}'',} \textit{ Phys. Lett. B}
  \textbf{ 605} (2005) 256,
  \href{http://dx.doi.org/10.1016/j.physletb.2004.11.019}{\doi{10.1016/j.physletb.2004.11.019}},
\href{http://www.arXiv.org/abs/hep-ex/0408005}{\texttt{arXiv:hep-ex/0408005}}.

\bibitem{fixed4}
\hrefCMSnoop {}{{CLAS} Collaboration, ``{Exclusive $\rho^{0}$ electroproduction
  on the proton at {CLAS}}'',} \textit{ Eur. Phys. J. A} \textbf{ 39} (2009) 5,
  \href{http://dx.doi.org/10.1140/epja/i2008-10683-5}{\doi{10.1140/epja/i2008-10683-5}},
\href{http://www.arXiv.org/abs/0807.3834}{\texttt{arXiv:0807.3834}}.

\end{thebibliography}\endgroup
